\documentclass[aps,amsfonts,preprint,nofootinbib,superscriptaddress,showkeys]{revtex4}

\usepackage{tikz}
\def\S2{\bar{S}}

\def\b{\beta}

\def\t{\theta}
\def\an{a_{n}}
\def\ta{\tilde{a}_n}
\def\and{a_{n}^\dagger}

\def\tab{\tilde{\bar{a}}_n}

\def\sn2d{\Sn2^\dagger}

\def\w{{\omega}}

\def\({\left(}
\def\){\right)}
\def\<{\left\langle}
\def\>{\right\rangle}

\def\ad{a^\dagger}
\newcommand\ee{\end{eqnarray}}      %eqnarray
\newcommand\be{\begin{eqnarray}}
\newcommand\ba{\begin{array}}           %array
\newcommand\ea{\end{array}}
\newcommand\eeq{\end{equation}}     %eqnarray
\newcommand\beq{\begin{equation}}

\usepackage{mathrsfs}
\usepackage{bbold}
\usepackage{graphicx}
\usepackage{xcolor}
\usepackage{caption}
\usepackage{bbold}
\usepackage{graphicx}

\begin{document}

\title{Entanglement Entropy from TFD Entropy Operator}

\author{M. Dias}
\email{mfedias@gmail.com}
\affiliation{Universidade Federal de S\~ao Paulo, Departamento de F\'isica, Rua S\~ao Nicolau 210, CEP: 09913-030, Diadema, SP, Brasil}

\author{Daniel L. Nedel}
\email{daniel.nedel@unila.edu.br}
\affiliation{Universidade Federal da Integra\c{c}\~{a}o Latino-Americana, Instituto Latino-Americano de Ci\^{e}ncias da Vida e da Natureza, Av. Tancredo Neves 6731 bloco 06, CEP: 85867-970, Foz do Igua\c{c}u, PR, Brasil}

\author{C. R. Senise Jr.}
\email{carlos.senise@unifesp.br}
\affiliation{Universidade Federal de S\~ao Paulo, Departamento de F\'isica, Rua S\~ao Nicolau 210, CEP: 09913-030, Diadema, SP, Brasil}
 

\begin{abstract}
In this work, a canonical method to compute entanglement entropy is proposed. We show that for two-dimensional conformal theories defined in a torus, a choice of moduli space allows the typical entropy operator of the TFD to provide the entanglement entropy of the degrees of freedom defined in a segment and their complement. In this procedure, it is not necessary to make an analytic continuation from the R\'enyi entropy and the von Neumann entanglement entropy is calculated directly from the expected value of an entanglement entropy operator. We also propose a model for the evolution of the entanglement entropy and show that it grows linearly with time.
\end{abstract}

\maketitle

%%%%%%%%%%%%%%%%%%%%%%%%%%%%%%%%%%%%%%%%%%%%%%%%%%%%%%%%%%%%%%%%%%%%%%%%%%%%%%%%%%%%%%%%%%%%%%%%%%%%%%%%%%%%%%%%%%%%%%%%%%%%%%%%%%%%%%%%%%%

\section{Introduction}

Entanglement is fundamental to quantum mechanics and it is one of the features that distinguishes it sharply from classical mechanics. The quantum entanglement plays an important role in quantum computations \cite{app} and the entanglement structure encoded in a many-body wavefunction provides important insights into the structure of the quantum state under consideration. The quantity that measures the quantum entanglement of a subset ${\mathcal H_{A}}$ of the Hilbert space $\mathcal H$ with the rest of this space is the entanglement entropy. Usually, the entanglement entropy is defined as the von Neumann entropy corresponding to the reduced density matrix  
\begin{equation}
S_E=-Tr \rho_A\ln\rho_A \ ,
\end{equation}
where the reduced density matrix $\rho_A$  of a subspace $\mathcal{A}$ of the Hilbert space $\mathcal H$ is obtained by tracing over the degrees of freedom of its complement $\tilde{\mathcal {A}}$. It is possible to use the reduced density matrix to define other types of entropies, such as R\'enyi entropies, determined by:
\begin{equation}
S^{(r)}_{A}=\frac{1}{1-r}\ln\rho_A^r \ ,
\end{equation}
where, in the canonical definition, $r\in\mathbb{Z}_+$, but an analytical continuation can be made to $r\in\mathbb{R}_+$, which is used to calculate the entanglement entropy from the limit
\begin{equation}
S_E= \displaystyle\lim_{r\rightarrow 1}S^{(r)}_{A} \ .
\end{equation}

In recent years, the entanglement entropy concept has been applied in several areas of theoretical physics: from condensed matter, where for some systems it was realized that the entanglement entropy is used as an order parameter \cite{Kitaev,Levin}, up to high energy physics, where entanglement has been playing an important role in the ADS/CFT conjecture and has helped to understand the thermodynamic entropy of black holes \cite{Maldacena,RT}. 

In general, the entanglement concept is better understood in discretized systems. To go into the realm of quantum field theory, we define the field $\phi(x)$, where $x$ is a set of spatial coordinates that describes the spatial location on a time-slice. Then, given a wave functional $\Psi[\phi(x)]$ for the instantaneous state of the system, the construction of the reduced density matrix can be realized in the same way as in the discrete case, as well as its associated entanglement entropy.

Although the wave functional $\Psi[\phi(x)]$ is useful to formally define the reduced density matrix in continuum QFTs, in general, in order to calculate the entanglement entropy for quantum fields, it is more useful to go directly to a path integral formulation. In this procedure, the entanglement entropy is not calculated directly. Actually, it is the R\'enyi entropy that is calculated and the entanglement entropy itself is obtained by means of an analytic continuation\footnote{For a review see \cite{takareview}.}. To do this calculation, it is necessary to define a contour in the functional integral which resembles the Schwinger-Keldysh (SK) formalism for real time finite temperature quantum field theory \cite{SK}. As usual, the contour used in the definition of the functional integral implies a doubling of the degrees of freedom.  

In despite of all the progress that have been made using the replica trick, a number of much more fundamental issues related to quantum field entanglement have not been completely addressed yet. The basic assumption that the Hilbert space factorizes into a direct product of degrees of freedom on $\mathcal{A}$ and its complement $\tilde{\mathcal{A}}$ fails in many cases, as pointed out in \cite{witten} . Even in the case that the Hilbert space factorizes, a direct calculation of the entanglement entropy requires a choice of boundary conditions at the edge of the chosen entangling region, which is in general non-physical. Furthermore, in relation to the temporal evolution of entropy, in general there are difficulties in implementing time translation in an unitary way \cite{Reznik:1995hy,Banks:1983by,Brustein:2006wp}.

Since the traditional way of calculating entanglement entropy is based on an analogy with the SK formalism for finite temperature field theory, it is natural to ask whether an alternative way could be found via an analogy with Thermo Field Dynamics (TFD) \cite{ume2,ume4}. Since TFD is an algebraic canonical formulation, such a method could provide a more rigorous way to obtain the entanglement entropy, without resort to the replica trick and euclidean path integral manipulations\footnote{TFD also has a functional representation \cite{KKMS}. The relationship between the perturbation scheme in TFD and the one in the SK formalism was first pointed out in \cite{Arimitsu,Marinaro}.}. The main goal of this work is to show that it is possible to construct a way to calculate the entanglement entropy using the  canonical TFD tools. 

Whereas in the SK formalism the duplication of the degrees of freedom is a consequence of the contour used in the functional integral, in the TFD formalism the duplication is defined from the beginning, in the context of the Tomita-Takesaki modular theory \cite{TT}. Thus, while in the SK formulation the time path fields represent the operator algebra in ordinary Fock space, in TFD the doublet fields are related to the GNS representation induced by KMS states. In this case, the cyclic (and separating) vector of the GNS triple is related to the TFD thermal vacuum and the equilibrium statistical average is written as expected values in this state. This proximity to the $C^{\star}$ approach to quantum statistical mechanics, in particular to the GNS construction, was laid down in \cite{ojima81}. It shows that TFD may be a natural framework for establishing a dictionary between a formal algebraic program to understand entanglement and a more physical picture,  especially within the holographic context, where algebraic treatment is necessary, as emphasized in \cite{kamal,Harlow:2016vwg}.

An important ingredient of the TFD formalism is the construction of an operator whose expected value in the thermal vacuum provides the thermodynamic entropy of the system: this is called the entropy operator. By construction, the entropy operator measures the entanglement between the original system and its copy; actually, the thermal vacuum is the state of maximum entanglement. This is not surprising, since the KMS state is a mixed state\footnote{For a review on the notion of pure states and mixtures in the $C^{\star}$-algebra context see Ref.~\cite{barata}.}. The central point that we raise here is that the TFD apparatus allows, within the GNS representation, a construction from a more general mixed state and not necessarily a thermal one. This is the central issue of this work.

We will show that we can generalize the TFD vacuum, such that it corresponds to a boundary state which is a more general KMS state, in the sense that it confines the fields in a temporal segment $\beta$. We apply the methodology to $c=1$ two-dimensional conformal theories with periodic and Neumann boundary conditions. A choice of torus moduli space allows the typical TFD entropy operator to provide the entanglement entropy of the degrees of freedom defined in a spatial segment and their complement. In this computation, the entanglement entropy is calculated directly and in a canonical way. Also, based on the non-equilibrium TFD, a model for the entanglement evolution is proposed. The model consists of a linear coupling between the original and auxiliary modes and the typical linear behavior of conformal theories is found in the small coupling limit. The entangled state at large times is not unitary equivalent to the initial state, which means that the system evolves to another representation of the operator algebra.

Since the strength of the methodology presented here lies in the proximity of the TFD formalism with the $C^{\star}$ approach to quantum fields, this work begins in Sec.~\ref{real} with a review of this approach, in order to show the conceptual differences and similarities between TFD and the SK formalism. In the sequence of this Section, we present a conformal transformation, in order to define the TFD thermal vacuum, generated by an entanglement entropy operator. The results for periodic boundary conditions are presented in Sec.~\ref{TFDsub} and for Neumann boundary conditions in Sec.~\ref{Neumann}. In Sec.~\ref{dissipative} we present a model for time-dependent entanglement. The conclusions are presented in Sec.~\ref{conc}.

%%%%%%%%%%%%%%%%%%%%%%%%%%%%%%%%%%%%%%%%%%%%%%%%%%%%%%%%%%%%%%%%%%%%%

\section{Real time Formalism and TFD} \label{real}

Since the usual method for calculating entanglement entropy is based on an analogy with the SK formalism to study quantum fields at finite temperature in real time, we will start this work with a general review of this formalism, in order to show some conceptual advantages that appear in the formulation of TFD. In particular, we will show the intimate relationship between TFD and the algebraic formulation of statistical mechanics. As will become clear in the following sections, it is precisely at this point that the main motivation of the present work lies.
  
Consider the Schr\"{o}dinger picture description of a general quantum mechanical system in a mixed state, described by a density matrix $\rho(t)$, which satisfy the quantum Liouville equation:
\begin{equation}
i\frac{\partial\rho(t)}{\partial t}=[H,\rho(t)] \ ,
\end{equation}
where $H$ is the hamiltonian of the system. For a general time-dependent hamiltonian, one can express the density matrix through the time evolution operator $U(t,t')$ as
\begin{equation}
\rho(t)= U(t,0)\rho(0)U^{\dagger}(t,0)=U(t,0)\rho(0)U(0,t) \ , \label{tddm}    
\end{equation}
\begin{equation}
i\frac{\partial U(t,t')}{\partial t}=H(t)U(t,t') \ .    
\end{equation}
	
Let us assume that $\rho(0)$ is the thermal equilibrium density matrix, at the equilibrium temperature $\beta^{-1}$:
\be
\rho(0)=\frac{e^{-\beta H_i}}{Tr e^{-\beta H_i}} \ . \label{roequi}
\ee
Without loss of generality, we suppose that the system is governed by the following hamiltonian:
\be
H(t)=\left\{\begin{array}{c}
       H_i \ \ \ , \ \mbox{for} \ {\it Re} \ t\leq0 \ , \\
       H(t) \ \ \ , \ \mbox{for} \ {\it Re} \ t\geq0 \ . \hspace{0,3cm}
       \end{array}\right.
\ee
Thus $\rho(0)$ can be written as
\be
\rho(0)= \frac{U(T-i\beta,T)}{Tr \ U(T-i\beta,T)} \ ,
\ee
where $T$ represents a large negative time, for which the hamiltonian is time-independent. Equation (\ref{tddm}) becomes: 
\be
\rho(t)= \frac{U(t,0)U(T-i\beta,T)U(0,t)}{Tr \ U(T-i\beta,T)} \ .
\ee
The time evolution of the expected value of any operator can be calculated in the usual way:
\be
\left\langle O\right\rangle_{\beta}(t)=Tr\rho(t)O \ ,
\ee
and using known properties of $U(t, t^{\prime})$ , the cyclic property of the trace and defining $T^{\prime}$ as a large positive time, we obtain:
\be
\left\langle O\right\rangle(t)= \frac{Tr \ U(T-i\beta,T) \ U(T,T^{\prime}) \ U(T^{\prime},t) \ O \ U(t,T)}{Tr \ U(T-i\beta,T)} \ .
\ee
This expression is the heart of the closed time path formalism. The system evolves from a large negative time $T$ to some time $t$ where an operator $O$ is inserted. Then, the system evolves from $t$ to a large positive time $T^{\prime}$ and so it evolves backward in time to $T$ and finally along an imaginary time. The contour $C$ in the complex $t$ plane is represented in Fig.~(\ref{Fig_t}).
\begin{center}
\includegraphics[scale=0.25]{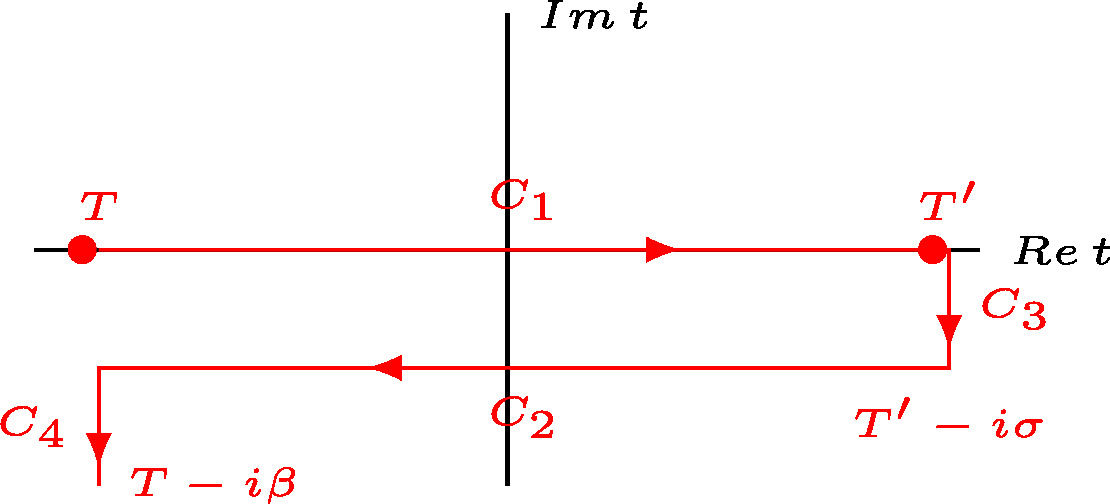}
\captionof{figure}{Contours in the complex $t$ plane.}
\label{Fig_t}
\end{center}
Note that the expected values should not depend on $\sigma$. 
  
We can take the limits $T\rightarrow -\infty$ and  $T'\rightarrow\infty$ and define the generating functional as:
\be
Z[J]=Tr \ U_{J}(T-i\beta,T) \ U_J(T,T') \ U_J(T',T) \ .
\ee
For a single scalar field in $D$ dimensions, the generating functional for the free theory limit can be written as
\be
Z_0[J]=N\exp\left\{-\frac{i}{2}\left[\int_{C_{12}}\int_{C_{12}}\!\!+\!\!\int_{C_{12}}\int_{C_{34}} \!\!+\!\!\int_{C_{34}}\int_{C_{12}}\!\!+\!\!\int_{C_{34}}\int_{C_{34}}\right]d^Dxd^Dx^{\prime}J(x)G(x-x')J(x')\right\} \ ,
\ee
where the integration contour has been split into four pieces and $C_{rs}= C_r\cup C_s$. In order to show that there is no contribution from the piece $C_3$, the Riemann-Lebesgue lemma \cite{kadiring} is used to show that 
\begin{equation}
\lim_{t\rightarrow \infty}\int_{C_{1}}\int_{C_{3}}d^{D}x \ d^{D}x^{\prime} \ J(x)G(x-x')J(x')=0 \ , \label{dec}
\end{equation}
if we have, in addiction, $\lim_{t\rightarrow\pm \infty}J(x)=0$. A similar argument holds for the pieces $C_2$ and $C_4$. Thus the generating functional factorizes into a contribution from $C_{12}$ and  $C_{34}$: $Z[J]=NZ_{12}[J]Z_{34}[J]$. For doing the calculations of the real time Green functions, the factor $Z_{34}[J]$ is constant and can be absorbed in the normalization constant $N$. So, all the dynamics is given by the contour $C_{12}$. This suggests that one can define a doublet of fields and sources as
\be
\phi_a=\left(\begin{array}{c}
             \phi_1 \\
             \phi_2
             \end{array}\right) \ \ \ , \ \ \ 
J_a=\left(\begin{array}{c}
             J_1 \\
             J_2
             \end{array}\right) \ ,
\ee
where $a=1,2$; $\phi_1, J_1 \in C_1$ and $\phi_2, J_2 \in C_2$. The generating functional can be written as
\be
Z[J_1,J_2]=N\exp\left\{-\frac{i}{2}\int d^{D}xd^{D}x'J^{a}(x)G_{ab}(x-x')J^{b}(x')\right\} \ ,	
\ee
and the Green's functions has a matrix structure such that, for $\sigma =\beta/2$:
\begin{equation}
iG_{ab}^{\beta}(x-x')=(-i)^2\frac{1}{Z}\left.\frac{\delta^2Z[J]}{\delta J^a(x)\delta J^b(x')}\right|_{J=0}=\mathbb {B}_k^{-1}(\beta) \ iG_{ab}(x-x') \ \mathbb {B}_k(\beta) \ , \label{prop}
\end{equation}
where  $G_{ab}(x-x')$ is the zero temperature propagator and $\mathbb {B}_k(\beta)$ is a unimodular matrix that carries all the thermal information. It will be written explicitly when we discuss TFD.
	
The form of the propagator, written in terms of a unimodular matrix that carries all the thermal information, is achieved almost by accident. In the TFD formulation, this structure, as well as the doubling of the degrees of freedom, appears naturally. Also, there is a class of theories for which the asymptotic conditions $\lim_{t\rightarrow\pm \infty}J(x)=0$ cannot be implemented and the generating functional cannot be written as $Z[J]=NZ_{12}[J]Z_{34}[J]$ \cite{muw,mat1,mat2}. In this case, it is necessary to use the TFD formalism. The factorization of the generating functional is also problematic in massless theories, where the infrared divergences make it impossible to use the Riemann-Lebesgue lemma \cite{land85}.  
	
Finally, given two Heisenberg operators, $A_H(t)$ and $B_H(t')$, where $A_H(t)= e^{iH_it}Ae^{-iH_it}$, and the same for $B_H(t')$, the equilibrium statistical average $\left\langle A(t)B(t')\right\rangle_{\beta}$ is written in terms of $\rho(0)$ defined in Eq.~(\ref{roequi}). By using the cyclic proprieties of the trace, we obtain:
\be
\left\langle A(t)B(t')\right\rangle_{\beta}= Tr\left[\rho(0)A_H(t)B_H(t')\right]= \left\langle A(t)B(t')\right\rangle_{\beta} \ . \label{KMS1}
\ee
Such relations are known as the Kubo-Martin-Schwinger (KMS) relations and are fundamental for the equilibrium statistical mechanics. Here they arise from the fact that we are evaluating a trace when the statistical average is calculated. However, in its most precise formulation, the KMS condition establishes an interplay between the proprieties of a given state and the time evolution of the operator algebra. One of the consequences of this interplay is the doubling of the degrees of freedom, as pointed out in the Tomita-Takesaki theory of von Neumann algebras. This is more natural in the TFD formalism, as we are going to show. An important characteristic of the state in question is that it is not pure. Therefore, the existence of this structure can be useful to study more general mixed states. In fact, we intend to show in this work that all the TFD technology can be useful to calculate quantities, such as entanglement entropy, in a more natural way, in the sense that it is most closely linked to the Tomita-Takesaki theory.  In order to see this fact, it is necessary to introduce some elements of the $C^{\star}$-algebra formulation of quantum statistical mechanics and its relation with TFD, which will be the subject of the next subsection. 	
	
%%%%%%%%%%%%%%%%%%%%%%%%%%%%%%%%%%%%%%%%%%%%%%%%%%%%%%%%%%%%%%%%%%%%%

\subsection{The $C^{\star}$-algebra approach and GNS construction} \label{CA}

In this subsection, the $C^{\star}$-algebra approach to quantum statistical mechanics will be briefly presented. A more complete review can be found in \cite{Fewster:2019ixc}. In the $C^{\star}$-algebra approach to many body quantum physics, the state $\psi$ is defined as a norm one positive linear functional on the operator algebra, rather than as a vector in a Hilbert space. Let us start with following definitions:

\noindent {\bf Definition 2.1.} A Banach $C^{\star}$-algebra $\mathcal{C}$ is a structure $(\mathbb{B},\mathbb{K},\cdot,+,\circ,\ast,\left\|. \right\|)$, consisting of:
\begin{itemize}
\item a set $\mathbb{B}$;

\item a field $\mathbb{K}=\mathbb{R},\mathbb{C}$;

\item a scalar product $\cdot$ : $\mathbb{K}\times \mathbb{B}\rightarrow\mathbb{B} $;

\item a vector sum $+$: $\mathbb{B}\times\mathbb{B}\rightarrow\mathbb{B}$;

\item an associative product $\circ$: $\mathbb{B}\times\mathbb{B}\rightarrow\mathbb{B}$;

\item an involution $\ast$:  $\mathbb{B}\rightarrow\mathbb{B}$;

\item a norm $\left\|. \right\|$ :  $\mathbb{B}\rightarrow \mathbb{R}$,
\end{itemize}
such that $(\mathbb{B},\mathbb{K},\cdot,+,\left\|. \right\|)$ is a Banach space over a field $\mathbb{K}$  and 
$(\mathbb{B},\mathbb{K},\cdot,+,\circ)$ is an associative algebra over the field $\mathbb{K}$, with an additional relation between the norm and the product:
\begin{equation}
\left\|A\circ B\right\|\leq A\circ B \ , \ \forall A,B \in \mathbb{B} \ .
\end{equation}
In the following, we are going to omit the symbol $\circ$ to indicate multiplication and use just $AB$ instead of $A\circ B$;

\noindent {\bf Definition 2.2.} An $\ast$ -representation $\pi$ of a $C^{\star}$-algebra $\mathcal{C}$ in a Hilbert space $\mathcal{H}$  is an $\ast$ -morphism of $\mathcal{C}$ in
the $C^{\star}$-algebra of the bounded operators in $\mathcal{H}$, $\pi: \mathcal{C} \rightarrow \mathcal{B}(\mathcal{H})$. A representation is said to be faithful if $\pi$ is injective;
 
\noindent{\bf Definition 2.3.} If $\mathcal{C}$ is a $C^{\star}$-algebra, an application $\phi: \mathcal{C} \rightarrow \mathbb{C}$ is said to be a linear functional if $\phi(\alpha a+\beta b)=\alpha\phi(a)+\beta\phi(b)$, for all $\alpha,\beta \in \mathbb{C}$ and $a,b\in \mathcal{C}$. A linear function is said to be positive if $\phi (a^{\star} a) \geq 0 $, for all $a\in \mathcal{C}$. A positive linear functional $\omega$ of an algebra $\mathcal{C}$ is said to be a state if it is normalized so that $\left\|\omega\right\|=1$.

In this work, an $\star$ -representation of a $C^{\star}$-algebra will be simply called a representation. For a $C^{\star}$-algebra that contains the unit (an unital algebra), the condition for normalizing the linear functional is equivalent to $\omega(\mathbb{1})=1$.
In order to regain the ordinary structure of quantum mechanics, the value $\omega(A)$ of the functional is interpreted as the expectation value of $A$ in the state $\omega$ and the $C^{\star}$-algebra is represented by a subset of the collection of all bounded operators $\mathcal{B}(\mathcal{H})$ on the Hilbert space $\mathcal{H}$. This is achieved by the  GNS construction, where GNS stands for Gelfand, Naimark and Segal \cite{GN,S}. The core of the GNS construction is the triple $(\mathcal{H}_{\omega}, \pi_\omega,
\Omega_{\omega})$, composed of the Hilbert space $\mathcal{H}_{\omega}$ , the $\pi_{\omega}$ representation and the vector $\Omega_{\omega}\in \mathcal{H}_\omega$. The triple $(\mathcal{H}_{\omega}, \pi_\omega, \Omega_{\omega})$ is
often called the GNS triple associated with the pair $(\mathcal{C}, \omega)$, composed of a $C^{\star}$-algebra  $\mathcal{C}$ and the state $\omega$ over $\mathcal{C}$. Once a state $\omega$ is chosen, the GNS construction of the Hilbert space $\mathcal{H}_{\omega}$  and the representation $\pi_{\omega}$ induced by $\omega$ proceeds as follows. Define the following closed linear subspace of $\mathcal{C}$ :
\begin{equation}
\mathcal{N}=\{n\in \mathcal{C} \ | \ \omega(n^{\ast}n)=0\} \ . \label{Ngns}
\end{equation}
Note that $\mathcal{N}=\mathcal{N}_1= \{n\in \mathcal{N} \ | \ \omega(b^{\ast}n)=0 \ , \ \forall b\in \mathcal{C}\}$. Furthermore, for all $n\in \mathcal{N}$ and $a\in \mathcal{C}$, we have that $an\in \mathcal{N}$. Thus, $\mathcal{N}$ is a left ideal of $\mathcal{C}$ .

Since $\mathcal{N}$ is a subspace of $\mathcal{C}$, we can build the quotient subspace $\mathcal{C} / \mathcal{N}$, formed by the equivalence classes: $[a]=\{ a+n \ , \ n \in {\cal N}\}$, $a \in \mathcal {C}$ . It possesses the null vector $[0]= [n \ , \ n\in \mathcal{N}]=\mathcal{N}$. 
Now we define, in $\mathcal {C} / \mathcal {N}$, the positive sesquilinear form, given by:
\begin{equation}
 \left\langle [a],[b]\right\rangle= \omega (a^{\ast}b) \ . \label {sesql}
\end{equation}
Note that if we replace $a$ by $a+n$, with $n\in\mathcal{N}$ we obtain:
\begin{equation}
\omega((a+n)^{\ast}b)=\omega(a^{\ast}b)+\omega(n^*b)=\omega(a^{\ast}b) \ .
\end{equation}
Thus, Eq.~(\ref{sesql}) does not depend on the representative taken in the classes. In fact, (\ref{sesql}) is a well defined scalar product. In particular, note that $\left\langle [a],[b]\right\rangle=\omega(a^{\ast}b)$  is zero if and only if $a\in\mathcal{N}$, in which case we would have $[a]=[0]$. With this procedure we have constructed a Hilbert space from the $C^{\star}$-algebra\footnote{In general, $\mathcal {C} / \mathcal {N}$ is not complete in relation to the norm induced by this scalar product, but we can consider its canonical completion $\widetilde{\mathcal {C} / \mathcal {N}}$ as the Hilbert space $\mathcal{H}_{\omega}$, which is complete.}. 

Let us now move on to the construction of the $\pi_\omega$ representation of the  algebra $\mathcal{C}$. For $a, z \in \mathcal {C}$, we define $\pi_{\omega}(a)$ in $\mathcal {C} / \mathcal {N}$ as follows: 
\begin{equation}
\pi_w[a]z=[az] \ . \label{piw}
\end{equation}
Note that if we replace $z$ by $z+n$, with $n \in {\cal N}$, we obtain $[az+an]=[az]$, since $an\in\mathcal{N}$. Thus, Eq.~(\ref{piw}) is well defined in the sense that it does not depend on the element $z$ taken in the class. Also,
\begin{eqnarray}
\pi_{\omega}(\alpha a+ \beta b)[z]& =& \alpha\pi_{\omega}(a)[z]+ \beta\pi_{\omega}[z] \ , \nonumber \\
\pi_{\omega}(a)\pi_{\omega}(b)[z]&=&\pi_{\omega}(ab)[z] \ ,
\end{eqnarray}
$\forall \alpha, \beta \in \mathbb{C}$ and $a,b \in \mathcal{C}$. It can be also shown that $\pi_{\omega}(a) $ is a bounded operator acting in $\mathcal{C} / \mathcal{N}$ and $\pi(a)^{\ast}=\pi(a^\dagger)$, where the dagger denotes the adjunct in $\mathcal {H}_{\omega}$. 

Now, let us show that there is a cyclical vector $\Omega_{w}$ that ``represents'' the state $\omega$ in $\mathcal{H}_{\omega}$, such that $\omega(a) =\left\langle \Omega_w,\pi_{\omega}\Omega_w\right\rangle$. If $\mathcal{C}$ is a Unital Algebra, we just define $\Omega_w$ as $\Omega_{\omega}:=[\mathbb{1}]$, so that
\begin{equation}
\left\langle\Omega_{\omega},\pi_{\omega}(a)\Omega_{\omega} \right\rangle =\left\langle[\mathbb{1},\pi_{\omega}(a)\mathbb{1} \right\rangle =\left\langle \mathbb{1},[a\mathbb{1}]\right\rangle  = \left\langle \mathbb{1},[a]\right\rangle =\omega(a) \ . \label{rep}
\end{equation}
Note that $\Omega$ is cyclical, because $\{\pi_{\omega}(a)\Omega_{\omega},a\in \mathcal{C}\} =\{[a],a\in \mathcal{C}\} = \mathcal{C}/\mathcal{N}$ and $\mathcal{C}/\mathcal{N}$ is obviously dense in $\mathcal{H}_\omega$.

Two representations $\pi_{\omega}$ and $\pi_{\omega'}$ are said to be unitary equivalent if there exists an isomorphism $U: \mathcal{H}_{\omega}\rightarrow\mathcal{H}_{\omega'}$
such that $\pi_{\omega}(a)=U\pi_{\omega'}(a)U^{-1}, \forall a\in \mathcal{C}$. So, the GNS triple is unique up to unitarity equivalence. The fact that there is no unitary equivalent representations $\pi_{\omega}$ induced by states $\omega$ helps to understand important physical situations, for example when symmetries are broken, where each phase is characterized by vacuum states that are unitary inequivalent. Also, as we are going to discuss in the next sections, two TFD thermal vacuum defined at different equilibrium temperatures are not unitary equivalent. This lies on the fact that vacuum states are represented by two states induced by non-unitary equivalent GNS constructions. So, each vacuum state induces a truly different representation of the operator algebra.  In the same direction, an important example  which is close-banded with this work is the time evolution of entanglement states \cite{nos}. In this case,  time translation cannot be unitarily implemented and the entanglement vacuum induces a different representation of the operator algebra  at each time. We will see this in sec.~\ref{dissipative}. An important theorem \cite{Emch,braro} states that any representation $\pi$ with a cyclic vector $\Omega$ satisfying $(\Omega,\pi(a)\Omega)=\left\langle \omega,a\right\rangle, \forall a\in \mathcal{C}$, is unitary equivalent to the GNS representation $\pi_{\omega}$.

Let us finish this subsection with an important relationship between the irreducibility of the GNS representation and the purity of the state. The collection of all states in a $C^{\star}$-algebra $\mathcal{C}$ is a convex set. In fact, if $\omega_1$ and $\omega_2$ are states in $\mathcal{C}$, then, for all $\lambda \in [0, 1]$,
$\omega=\lambda\omega_1 + (1 - \lambda) \omega_2$ is also a state in $\mathcal{C}$. The state $\omega$ is said to be a mixed state if there are $\lambda \in  (0, 1)$ and $\omega_1, \omega_2$, with $\omega_1 \neq \omega$ and $\omega_2\neq\omega$, such that $\omega=\lambda\omega_1 + (1 - \lambda) \omega_2$. A state $\omega$ is said to be a pure state if it is not a mixed state.

The GNS representation $(\mathcal{H}_{\omega}, \pi_\omega,
\Omega_{\omega})$ is irreducible if and only if $\omega$ is a pure state on $\mathcal{C}$, and hence it cannot be written as a mixture of other states; alternatively, any mixed state on $\mathcal{C}$ induces a reducible GNS representation of the algebra. Finally, if the representation is irreducible, and thus $\psi$ is pure, the commutant  $\pi(\mathcal{C})'$ is trivial (i.e. equal to $\{\alpha \mathbb{1}\}$, $\alpha \in \mathbb{C}$). Otherwise, if $\pi(\mathcal{C})'\neq \alpha \mathbb{1}$, the state is a mixed one.

%%%%%%%%%%%%%%%%%%%%%%%%%%%%%%%%%%%%%%%%%%%%%%%%%%%%%%%%%%%%%%%%%%%%%%%%

\subsection{Tomita-Takesaki theorems and KMS states} \label{TomTa}

The Tomita-Takesaki modular theory was responsible for significant advances in Operator Algebra and its applications in Theoretical Physics. The relationship between the modular theory of Tomita and Takesaki and Gibbs' states was first pointed out by Haag, Hugenholtz and Winnink in Ref.~\cite{hhw}. In order to introduce the basic concepts that help us to understand the TFD formalism, we need some rather unusual concept of anti-linear operators between Hilbert spaces, associated to anti-isomorphisms between $C^{\star}$-algebras. A linear bijection $\phi : \mathcal{C}_1 \rightarrow \mathcal{C}_2$ between $C^{\star}$-algebras is an anti-isomorphism if it preserves involution and for $a,b \in \mathcal{C}_1$, $\phi(ab)=\phi(b)\phi(a)$. In this case, $\mathcal{C}_1$ and $\mathcal{C}_2$ are anti-isomorphic. The Tomita-Takesaki theorem constructs a family of anti-isomorphisms between $\mathcal{C}_1$ and $\mathcal{C}_2$ which preserves involution. 

Let us assume that the $C^{\star}$-algebra $\mathcal{C}$ is represented by bounded operators on a Hilbert space $\mathcal{H}$ by means of the GNS triple and that $A\Omega=0$ implies $A=0$, for $A \in \mathcal{M}$, where $\mathcal{M}$ is the enveloping von Neumann algebra of $\pi(\mathcal C)$. Thus, $\Omega$ is a cyclic and separating vector. The antilinear maping $S: \mathcal{H}\rightarrow \mathcal{H}$ is defined by $S:A\Omega=A^{\dagger}\Omega$. From the separating property of $\Omega$, $S$ is injective, while the fact that $\Omega$ is cyclic yields surjectivity. These properties ensure that $S$ is an isomorphism of vector spaces.

Consider the polar decomposition of the Tomita-Takesaki operator $S$:
\begin{equation}
S = J\Delta^{1/2} = \Delta^{-1/2}J \ ,
\end{equation}
where $J$ is an anti-linear unitary operator, $J^{\ast}=J^{-1}$. The operator $\Delta S^*S$ is unique positive and self-adjoint ($(\Delta \Omega, \Omega) \geq 0$ ). The operator $\Delta$ is called the {\it modular operator} and $J$ is the {\it modular conjugation} (or {\it modular involution}) associated with the pair $(\mathcal{M},\Omega)$. One can also define the self-adjoint operator $H_{\Delta}=\ln\Delta$,  known as {\it modular hamiltonian}. Note that $J\Omega = \Omega = \Delta\Omega$ and $J^2=1$, which are the main properties of $\Delta$ and $J$.

The main theorem of the modular Tomita-Takesaki theory establishes a one to one correspondence between $\mathcal{M}$ and its commutant $\mathcal{M}'$, and gives rise to the doubled degrees of freedom, in that $\mathcal{M}'$ is a copy of $\mathcal{M}$:
\begin{equation}
J\mathcal{M}J= \mathcal{M}' \ \ , \ \ \Delta^{-it/\beta} \mathcal{M}\Delta^{it/\beta} = \mathcal{M} \ .\label{TTT}
\end{equation}
Note that $\Delta^{it/\beta}$ induces a one-parameter automorphism group $\{ \sigma_t \}$ of $\mathcal{M}$ by
\begin{equation}
\sigma_t(A) = \Delta^{it/\beta} A \Delta^{it/\beta} \quad , \quad A \in \mathcal{M} \ . \label{autm}
\end{equation}
This is the so called {\it modular automorphism group} of $\mathcal{M}$
(relative to $\Omega$). It leaves invariant the faithful normal state on $\mathcal{M}$ induced by $\Omega$: $\omega(\sigma_t(A)) = \omega(A)$, for all  $A \in \mathcal{M} $ and $t \in \mathbb{R}$.   The properties of the $\Delta$ operator allow us to identify $\sigma_t$ with the time evolution in the representation space. We also have
\begin{equation}
\left(\Omega,A\sigma_{t+i\beta}[B]\Omega\right)=(\Omega,\sigma_t[B]A\Omega) \ .
\end{equation}
This is exactly the KMS condition and it is a consequence of the existence of a cyclic and separating vector $\Omega$ in $\mathcal{H}$. This vector will be identified with the thermal vacuum in the TFD formalism.

%%%%%%%%%%%%%%%%%%%%%%%%%%%%%%%%%%%%%%%%%%%%%%%%%%%%%%%%%%%%%%%%%%%%%%%

\subsection{Explicit representation and TFD} \label{exp}

We will now construct a physical image from the abstract ideas discussed previously. For a Hilbert space $\mathcal{H}$, the set $\mathcal{B}(\mathcal{H})$ of the bounded linear operators
acting in $\mathcal{H}$ is a $C^{\star}$-algebra. For the case where $\mathcal{H}$ is the space of finite dimension $\mathbb{C}^n$, $ \mathcal{B}(\mathcal{H})$ coincides with the algebra
$Mat(\mathbb{C}, n)$ of $n \times n$ matrices with complex entries. $Mat(\mathbb{C}, n)$ is a vector space and we can define a scalar product given by $\left\langle A,B\right\rangle= Tr(A^{\dagger}B)$, for all $A,B \in Mat (\mathbb{C}, n)$.

A matrix $\rho \in Mat(\mathbb{C}, n)$ that is self-adjoint, positive and satisfies $Tr\rho = 1$ is said to be a density matrix, such that every state $\omega \in \mathcal{B}(H)=Mat(\mathbb{C}, n) $ is written as
\begin{equation}
\omega(A)= Tr(\rho A) \ .
\end{equation}
This allows one to identify the set of states in $Mat(\mathbb{C}, n)$ with the set of density matrices in $Mat(\mathbb{C}, n)$. 

We now move to the GNS construction. For the sake of simplicity, let us concentrate on the density matrices $\rho$ that are invertible \footnote{For a general construction we need first to construct the set $\mathcal{N}$ defined in a general way in (\ref{Ngns}). For this case, $\mathcal{N}$ is defined by $\mathcal{N}=\{N\in Mat(\mathbb{C}, n) \ | \ N\rho^{1/2}=0\}$, which is equivalent to $\mathcal{N}=\{N\in Mat(\mathbb{C}, n) \ | \ Ker N\subset Ran(\rho^{1/2})\}$. If $P_{\rho}$ is the orthogonal projector over $Ran(\rho^{1/2})$, the space $Mat(\mathbb{C}, n)/\mathcal{N}$ is $Mat(\mathbb{C}, n)/\mathcal{N}\equiv \{AP_{\rho} \ | \ A\in Mat (\mathbb{C}, n) \} $, such that we can define in $Mat(\mathbb{C}, n)/\mathcal{N}$ a well defined scalar product by $\left\langle AP_{\rho},BP_{\rho}\right\rangle_{\rho}=Tr(\rho P_{\rho}A^{\ast}BP_{\rho})=Tr(P_{\rho}\rho AP_{\rho}B)=Tr(\rho A^{\ast}B)=\omega_{\rho}(A^{\ast}N)$.}. Obviously, it is not quite general, but since in this work  we are interested in Gibbs' states, this will not be a problem. For a Gibbs' state $\omega_{\beta}$, we have $\rho_{\beta}=\frac{e^{-\beta H}}{Tr e^{-\beta H}}$, so that $\left\langle \omega_{\beta};A\right\rangle$ is the expectation value of $A$ at the equilibrium temperature $\beta^{-1}$. In this case, we use $\Omega_\rho=\rho_{\beta}^{1/2}$ as the cyclic vector of the GNS triple. The representation $\pi_{\omega}$ is defined by
\begin{equation}
\pi_{\omega}(A)B=AB \ , \label{lr}
\end{equation}
for all $A, B \in Mat(\mathbb{C}, n)$. For $\Omega_\rho=\rho_{\beta}^{1/2}$, Eq.~(\ref{rep}) becomes 
\begin{equation}
(\Omega,\pi_{\omega}(A)\Omega)= Tr\rho\pi_{\omega}A=\left\langle \omega,A\right\rangle \ . \label{me} 
\end{equation}

In order to use the Tomita-Takesaki theory, let us construct the commutant $\pi_{\omega}(\mathcal{C})'$ of $\pi_{\omega}(\mathcal{C})$, as well as the operators $J$ and $\Delta$. Based on (\ref{lr}), we define the following antilinear representation $\nu_\omega$:
\begin{equation}
\nu_{\omega}(A)B=BA^{\dagger} \ . \label{alr}
\end{equation}
If we define $J$ by
\begin{equation}
JB=B^{\dagger} \ , \label{tom}
\end{equation}
we obtain
\begin{equation}
J\pi_\omega(A)JB=JAB^{\dagger}=BA^{\dagger}=\nu_{\omega}(A)B \ ,
\end{equation}
and it is easy to see that $\nu_{\omega}(A)\pi_{\omega}(B)=\pi_{\omega}(B)\nu_{\omega}(A)$, for all $A,B\in  Mat(\mathbb{C}, n)$. Thus, we have in fact constructed an antilinear map between $\pi_{\omega}(\mathcal{C})$ and its commutant. 

Now we construct the modular operator $\Delta$, such that $\Delta \Omega=\Omega$. By using the so-called Liouville operator:
\begin{equation}
\hat{H}=\pi_{\omega}(H)-\nu_{\omega}(H) \ , \label{LO}
\end{equation}
we define $\Delta=e^{\hat{H}}$. Since $\pi_{\omega}(H)$ and $\nu_{\omega}(H)$ commute, it is easy to show that $\Delta \Omega=\Omega$, for $\Omega=\rho^{1/2}$, which is equivalent to
\begin{equation}
\hat{H}\Omega=0 \ . \label{homeg}
\end{equation}

The one-parameter automorphism group $\{ \sigma_t \}$ of $\mathcal{M}$ is now defined by
\begin{equation}
\sigma_{t}[\pi_{\omega}(A)]B=\Delta^{-it/\beta}\pi_{\omega}(A)\Delta^{it/\beta}B = e^{-iH}Ae^{iHt}B \ ,
\end{equation}
and the KMS condition follows as usual. In this manner, $\Delta^{-it/\beta}$ is the time evolution operator in the full space $\pi_{\omega}(\mathcal{C})\cup\nu_{\omega}(\mathcal{C})$. 

We finish this subsection with a very important result for the development of this work. It is clear that the commutant satisfies $\nu_{\omega}(\mathcal{C})\neq \mathbb{C}\mathbb{1}$, and so the Gibbs' state $\omega_\beta$ is a mixed state. The TFD formalism is based on this particular state. In the present work we are going to show that the TFD technology (at least in conformal theories) can be used to study more general mixed states.

%%%%%%%%%%%%%%%%%%%%%%%%%%%%%%%%%%%%%%%%%%%%%%%%%%%%%%%%%%%%%%%%%%%%%%%%

\subsection{TFD} \label{TFDrev}

TFD is a canonical formalism which enables one to study finite temperature field theory in real time. As showed by Ogima \cite{ojima81}, TFD can be described entirely in terms of the algebraic formalism described earlier. 

Let us just establish a dictionary between the algebraic terms of the previous section and the TFD terms, which will be the language used  from this point on. We are going to work only with bosonic fields and the generalization for fermionic ones is straightforward. The operator algebra will consist of the operator fields $\phi(x)$. Considering the GNS triple, the representative $\pi_{\omega}$ will be just $\phi(x)$. In the equilibrium TFD formalism at the temperature $\beta^{-1} $, the cyclic (and separating) vector $\Omega$ is the ket denoted by $\left|0(\beta)\right\rangle$, called the thermal vacuum. The modular conjugate of $\phi(x)$ is called the tilde field, denoted by $\tilde{\phi}(x)$, so that 
\begin{equation}
J\phi(x)J=\tilde{\phi}(x) \quad , \quad [\phi(x),\tilde{\phi}(x)]=0 \ .
\end{equation}
Equation (\ref{tom}) implies, for any operator $A(\phi)$ and $c\in \mathbb{C}$, the so-called tilde conjugation rules:
\begin{eqnarray}
(A_{i}(\phi)A_{j}(\phi))\widetilde{}&=&\widetilde{A}_{i}(\phi)\widetilde{A}_{j}(\phi) \ , \nonumber \\
(cA_{i}(\phi)+A_{j}(\phi))\widetilde{}&=&c^{\ast}\widetilde{A}_{i}(\phi)+\widetilde{A}_{j}(\phi) \ , \nonumber \\
(A_{i}^{\dagger }(\phi))\widetilde{}&=&(\widetilde{A}_{i})^{\dagger }(\phi) , \:\: (\widetilde{A}_{i})\widetilde{}(\phi) =A_{i}(\phi) \ , \label{til}
\end{eqnarray}
while Eq.~(\ref{me}) allows us to interpret the statistical average of an operator $A(\phi)$ as the expected value of $A(\phi)$ in the thermal vacuum: 
\begin{equation}
\left\langle A(\phi)\right\rangle_{\beta}=\left\langle 0(\beta )\left|A(\phi)\right|0(\beta )\right\rangle \ .
\end{equation}
Note that $J\Omega=0$ implies that the vacuum is tilde invariant. Time translations in the total Hilbert space ${\cal H}_{T}={\cal H}\otimes\widetilde{{\cal H}}$ is given, from (\ref{LO}), by $\hat{H}=H-\tilde{H}$, and (\ref{homeg}) implies that the thermal vacuum is annihilated by $\hat{H}$. By using the polar representation of the Tomita-Takesaki operator, we obtain the so-called thermal state condition:
\begin{equation}
e^{\beta H/2}\tilde{\phi}(x)\left|0(\beta)\right\rangle=\phi^{\dagger}(x)\left|0(\beta)\right\rangle \ . \label{tscg}
\end{equation}

Note that all the properties presented so far do not depend on any particular characteristic of the bosonic field-theoretic system. Thus, TFD is just a physical representation of the algebraic ideas previously discussed. This is the strength of the formalism.

Suppose that the hamiltonian  has a discrete spectrum and that it is possible to find a base $\left|\psi_n\right\rangle$ where the hamiltonian is diagonal. In this basis, $\left|0(\beta)\right\rangle$ becomes
\begin{eqnarray}
\left|0(\beta)\right\rangle&=& \frac{1}{Z^{1/2}}\sum_{n}e^{-\frac{\beta E_n}{2}}|\psi_n,\tilde{\psi_n}\rangle \nonumber \\
&=& e^{-\frac{S}{2}} \ I \qquad , \qquad I=\sum_n|\psi_n,\tilde{\psi_n}\rangle \ ,
\end{eqnarray}
where $Z =Tr e^{-\beta H}$ and
\begin{equation}
S= H-\ln Z \label{entgeral}
\end{equation}
is an operator whose expected value in $\left|0(\beta)\right\rangle$  provides the system's thermodynamic entropy. It plays a central role in this work, as will become clear below.

%%%%%%%%%%%%%%%%%%%%%%%%%%%%%%%%%%%%%%%%%%%%%%%%%%%%%%%%%%%%%%%%%%%%%%%

\subsection{The free theory limit} \label{free}

Let us turn our attention to free theories. From this point on, we will describe the formalism in a general way, so that the thermal vacuum can be seen as a more general mixed state, not necessarily related to the Gibbs' state. 

The first construction of TFD consists in defining a Bogoliubov transformation, using only one generator that produces an unitary transformation and preserves the tilde conjugation rules. A major generalization can be done and one can find a set of generators that maintains the thermal nature of the transformation. The set of generators forms an oscillator representation of the $SU(1,1)$ group for bosons and of the $SU(2)$ group for fermions \cite{ume1,ChuUme}; these are related to the Gaussian structure of the thermal vacuum. The $SU(1,1)$ formulation allows one to write the Bogoliubov operator in terms of complex parameters, which is essential for the results presented in the following subsection. 

Let us explore the $SU(1,1)$ unitary formulation for a set of bosonic oscillators. The $SU(1,1)$ Bogoliubov generators are\footnote{Actually, there are two ways to generalize the TFD Bogoliubov generator. In one case, the tilde conjugation rules are preserved but the transformation, even in a finite volume limit, is non-unitary. This construction was largely applied and its connection with other thermal field theories is clear, as one can see, for example, in Refs. \cite{ume1} and \cite{Hen}. In the other case, the transformation is unitary in a finite volume limit, but the tilde conjugation rules are not preserved, which would cause an ambiguity in the choice of the thermal vacuum. In Ref. \cite{gadcris}, the ambiguity related to the thermal vacuum choice was shown to be just apparent and the so-called general unitary $SU(1,1)$ TFD formulation was presented. Such a formulation considers a transformed Tomita-Takesaki modular operator. As a consequence, the tilde conjugation rules were redefined in the transformed space and named breve conjugation rules. In the application of  Ref. \cite{adsnos}, it is shown that the $SU(1,1)$ parameters are related to the boundary conditions of the string worldsheet fields in AdS black holes.}
\begin{equation}
G=\sum_{k=1}\left[\gamma_{1_{k}}\widetilde{a}_{k}^{\dagger}\cdot a^{\dagger }_{k}-\gamma_{2_{k}}a_{k}\cdot\widetilde{a}_{k}+\gamma_{3_{k}}\left(a^{\dagger }_{k}\cdot a_{k}+\widetilde{a}_{k}\cdot\widetilde{a}^{\dagger }_{k}\right)\right] \ , \label{ge}
\end{equation}
where $\gamma_{1_{k}}$, $\gamma_{2_{k}}$ and $\gamma_{3_{k}}$ are free parameters of the Bogoliubov transformation, and the thermal vacuum is
\begin{eqnarray}
\left|0(\t)\right\rangle&=&\exp(-iG)\left.\left|0\right\rangle\!\right\rangle \ .
\label{res1a}
\end{eqnarray}
In Eq. (\ref{res1a}), $\theta$ represents all the $SU(1,1)$ parameters. This generator carries out an unitary and canonical transformation such that the creation and annihilation operators transform according to
\begin{eqnarray}
\left(\begin{array}{c}
      a_{k}(\theta) \\
      \tilde{a}^{\dagger}_{k}(\theta)
      \end{array}
\right)&=&e^{-i{\bf G}}\left(\begin{array}{c}
                             a_{k} \\
                             \tilde{a}^{\dagger }_{k}
                             \end{array}
                        \right)e^{i{\bf G}}={\mathbb B}_{k}\left(\begin{array}{c}
                                    a_{k} \\
                                    \tilde{a}^{\dagger }_{k}
                                    \end{array}
                                    \right) \ ,%\label{tbt}
\nonumber\\
\left(\begin{array}{cc}
      a^{\dagger}_{k}(\theta) & -\tilde{a}_{k}(\theta)
      \end{array}
\right)&=&\left(\begin{array}{cc}
                a^{\dagger }_{k} & -\tilde{a}_{k}
                \end{array}
                \right){\mathbb B}^{-1}_{k} \ , \label{tbti}
\end{eqnarray}
where the matrix transformation is given by
\begin{eqnarray}
{\mathbb B}_{k}=\left(\begin{array}{cc}
                           \mathfrak{u}_{k} & \mathfrak{v}_{k} \\
                           \mathfrak{v}^{*}_{k} & \mathfrak{u}^{*}_{k}
                           \end{array}
                     \right) \ ,
\qquad|\mathfrak{u}_{k}|^{2}-|\mathfrak{v}_{k}|^{2}=1 \ , \label{tbm1}
\end{eqnarray}
with elements \cite{gad1}
\begin{equation}
\mathfrak{u}_{k}=\cosh\left(i\Gamma_{k} \right)+\frac{\gamma_{3_{k}}}{\Gamma_{k}}\sinh\left(i\Gamma_{k}\right) \ ,
\qquad
\mathfrak{v}_{k}=-\frac{\gamma_{1_{k}}}{\Gamma_{k}}\sinh\left(i\Gamma_{k}\right) \ , \label{uvexp2}
\end{equation}
and $\Gamma_{k}$ is defined by the following relation:
\begin{equation}
\Gamma^{2}_{k}=\gamma_{1_{k}}\gamma_{2_{k}}+\gamma_{3_{k}}^{2} \ . \label{Gadef}
\end{equation}
 
A quite convenient way to write the Bogoliubov transformation matrix (\ref{tbm1}) arises if we make the polar decomposition $\mathfrak{u}_{k}=|\mathfrak{u}_{k}|e^{i\varphi_{k}}$, $\mathfrak{v}_{k}=|\mathfrak{v}_{k}|e^{i\phi_{k}}$, and rewrite the matrix elements in terms of the new parameters:
\begin{equation}
f_{k}=\frac{|\mathfrak{v}_{k}|^{2}}{|\mathfrak{u}_{k}|^{2}} \ ,
\quad
\alpha_{k}=\frac{\log(\frac{\mathfrak{v}_{k}}{\mathfrak{u}_{k}})}{\log(f_{k})}=\frac{1}{2}+i\frac{(\phi_{k}-\varphi_{k})}{\log(f_{k})} \ ,
\quad
s_{k}=i\varphi_{k}=\frac{1}{2}\log\left(\frac{\mathfrak{u}_{k}}{\mathfrak{u}_{k}^{*}}\right) \ . \label{sfap}
\end{equation}
In fact, with these steps we can present the Bogoliubov matrix as \cite{ume1,Hen}
\begin{equation}
{\mathbb B}_{k}=\frac{1}{\sqrt{1-f_{k}}}\left(\begin{array}{cc}
                                                   e^{s_{k}} & -f_{k}^{\alpha_{k}}e^{s_{k}} \\
                                                   -f_{k}^{\alpha_{k}^{*}}e^{-s_{k}} & e^{-s_{k}}
                                                   \end{array}
                                             \right) \ , \label{sfapM}
\end{equation}
with $\alpha_{k}+\alpha^{*}_{k}=1$. Since $a(\theta)$ annihilates the thermal vacuum, we obtain the state condition:
\begin{equation}
\left[a_k-f_k^ {\alpha} \ \widetilde{a}_{k}^{\dagger}\right]|0(\theta)\rangle=0 \ .
\label{thermo2}
\end{equation}
Therefore, the thermal vacuum is in fact a boundary state. The expected value of the number operator gives the distribution:
\begin{equation} 
N_k(\t)=\left\langle0(\t)\right|a_k^{\dagger}a_k\left|0(\t)\right\rangle= \frac{f_{k}}{1-f_{k}}\ .
\label{distri1}
\end{equation}

At the equilibrium, the $SU(1,1)$ parameters are fixed as functions of the temperature by minimizing a thermodynamic potential. In order to do this, let us define a very important operator, the entropy operator \cite{gad2}:
\begin{equation}
K=-\displaystyle{\sum_{k=1}}\left[a_{k}^{\dagger}\cdot a_{k}\log\left(\frac{\gamma_{1_{k}}\gamma_{2_{k}}}{\Gamma_{k}^{2}}\sinh^{2}\left(i\Gamma_{k}\right)\right)-a_{k}\cdot a_{k}^{\dagger}\log\left(1+\frac{\gamma_{1_{k}}\gamma_{2_{k}}}{\Gamma_{k}^{2}}\sinh^{2}\left(i\Gamma_{k}\right)\right)\right] \ . \label{kag}
\end{equation}
It was shown in \cite{gad2} that the thermal vacuum can also be generated by the action of this  entropy operator. In a thermal equilibrium situation, its expected value provides the thermodynamic entropy, which has its origin in coarse graining, and is not related to entanglement. However, one can show that, in general, $K$ measures the entanglement between the original system and its copy. To make this statement clear, it can be shown that the thermal vacuum can be written as follows:
\begin{equation}
\left|0(\theta)\right\rangle=\displaystyle{\sum_{n}}\sqrt{{\cal W}_{n}(\theta_n)}\left|n,\tilde{n}\right\rangle \ ,
\end{equation}
where
\begin{equation}
{\cal W}_{n}(\t)=\prod_{k}\left(\frac{|\mathfrak{v}_{k}|^{2n_{k}}}{\left(|\mathfrak{u}_{k}|^2\right)^{n_{k}+1}}\right) \ .
\end{equation}

Now it is possible to make an important connection with the density operator. For the sake of simplicity, we are going to write the tilde fields as $B$ fields. Considering the so-called extended density matrix
\begin{equation}
\rho=\left|0(\theta)\right\rangle\left\langle0(\theta)\right| \ .
\end{equation}
If we trace over the tilde fields (now the $B$ fields), the reduced density operator is
\begin{eqnarray}
\rho_{A}&=&Tr_{B}\left[\left|0(\theta)\right\rangle\left\langle0(\theta)\right|\right] \nonumber\\
&=&\sum_{n}{\cal W}_{n}(\theta)\left|n\right\rangle\left\langle n\right| \ ,
\end{eqnarray}
and the entropy is written as
\begin{equation}
{\cal S}_{A}(\theta)=\left\langle0\left(\theta\right)\left|K _{A}\right|0\left(\theta\right)\right\rangle=\sum_{n}{\cal W}_{n}(\theta)\ln{\cal W}_{n}(\theta) \ . \label{entA-t}
\end{equation}
This makes clear the fact that the expected value of the entropy operator is the entanglement entropy relative to tracing over the $B$ degrees of freedom. However, the Hilbert space is not geometrically partitioned here - the division of the system into subsystems $A$ and $B$ does not follow the more traditional geometric delimitation. The entanglement is related to the boundary state solution of Eq.~(\ref{thermo2}). 

Note that, up to this point, the $SU(1,1)$ parameters are quite general and can be time-dependent. Now we are going to fix them. In TFD, the expected value of the original hamiltonian is interpreted as the thermal energy. The following potential is then defined:
\begin{equation}
F=\left\langle0\left(\theta\right)\left|H\right|0\left(\theta\right)\right\rangle-\frac{1}{\b}\left\langle0\left(\theta\right)\left|K _{A}\right|0\left(\theta\right)\right\rangle \ .
\end{equation}
Now, for a set of oscillators with frequency $\omega_k$, by minimizing $F$ in relation to the Bogoliubov parameters, we find
\begin{equation}
N_k=\frac{f_{k}}{1-f_{k}} \ \rightarrow \ N_k^{\beta}=\frac{1}{e^{\b\omega_k}-1}  \ , \label{terdistri}
\end{equation}
which is the usual thermodynamic distribution for a bosonic system in equilibrium at temperature $\beta^{-1}$. Now, the entropy operator provides the thermodynamic entropy and $F$ is the free energy. In this case, the free thermal propagator is
\begin{equation}
G_{ab}(x,y)=\left\langle 0(\beta)|T\left[\phi_a(x)\phi_b(y)\right]|0(\beta)\right\rangle=\left\langle 0 |\mathbb {B}_k(\beta)^{-1}T\left[\phi_a(x)\phi_b(y)\right]
\mathbb {B}_k(\beta)|0\right\rangle \ ,
\end{equation}
where the matrix $\mathbb{B}_k(\beta)$ is the Bogoluibov matrix evaluated with $N_k^{\beta}$ defined in (\ref{terdistri}), and it is exactly the matrix that appears in (\ref{prop}).  

We have shown that generally, in the TFD framework, the same operator that measures the entanglement between two sets of fields provides, for a given choice of the entanglement parameters, the thermodynamic entropy. This is quite amazing and fits with what we learn from string theory. In particular, in Ref.~\cite{Brustein}, it is shown that the entropy resulting from the counting of microstates of non-extreme black holes can be interpreted as arising from entanglement in the dual description. 

The entropy operator is a function of the distribution and it is quite general. We will show in the next section that we can choose the distribution such that the entropy operator corresponds to a spatial entanglement entropy, that is, a measure of entanglement in a spatially partitioned Hilbert space. Based on the GNS construction, this is not a surprise, since the TFD thermal vacuum is related to a cyclic and separating vector, which is the Gibbs' state. The TFD formalism allows one to describe a more general vacuum, related to a more general cyclic and separating vector.

%%%%%%%%%%%%%%%%%%%%%%%%%%%%%%%%%%%%%%%%%%%%%%%%%%%%%%%%%%%%%%%%%%%%%

\section{Entropy operator in conformal theories defined in a torus} \label{TFDsub}

It was shown in \cite{torus} that, for a two-dimensional conformal theory with central charge $c$, it is possible to use Eq. (\ref{thermo2}) to define a boundary state that represents a torus with moduli space $\tau=\lambda+i\frac{\beta}{2\pi}$, where $\lambda$ is a Lagrange multiplier associated to a gauge fixing. More precisely, the $SU(1,1)$ parameters are related to the torus moduli space and the thermal state condition defines a torus boundary state. In this particular construction, the moduli space is defined in such a way that the torus boundary space corresponds to the TFD thermal vacuum. However, as discussed in the previous section, the TFD formalism allows a more general choice for the parameters of the Bogoliubov transformation, which actually defines the boundary state. In this section, we will combine this fact with conformal invariance to define the more general torus moduli space, such that the expected value of the entropy operator is in fact the entanglement between the degrees of freedom of the theory defined in an interval and its complement. The inspiration for this issue is the pioneering work of \cite{wil}, which calculates the entanglement entropy from the torus partition function. From this point on, we will work with $c=1$.

The torus boundary state is derived from gluing conditions between the original field $\phi(x,t)$, defined in a cylinder, and an auxiliary field $\widetilde{\phi}(\tilde{x},\tilde{t})$, defined in another cylinder. We are just gluing together the end of the original cylinder with the origin of the tilde cylinder, and vice-versa. In order to define the cylinders, the $t$ parameter runs from $t$ to $t+\b/2$ in imaginary time and $\tilde{t}$ runs in the reverse direction. This procedure will confine the fields to a restricted region $\beta$ in the euclidean time. Also, before gluing, the identification $\tilde{x}=x-\pi\lambda$ must be done in order to take into account the Dehn twist in one cycle of the torus. The two parameters of the resulting torus moduli space will be related to $\b$ and $\lambda$. We can think of the fields as living in the two sides of a bifurcated Killing horizon, and admit a plane wave expansion. These considerations can be written as follows:
\begin{eqnarray}
\phi(t,x)-\widetilde{\phi}(-t-\frac{\b}{2},x-\lambda\pi)&=&0 \ , \nonumber\\
\phi(-\tilde{t}-\frac{\beta}{2},\tilde{x}+\lambda\pi)-\widetilde{\phi}(\tilde{t},\tilde{x})&=&0 \ . \label{ident}
\end{eqnarray}
Expanding $\phi(t,x)$ and $\widetilde{\phi}\(t,x\)$ in modes and assuming periodic boundary conditions, we obtain
\begin{equation}
\phi(t,x)=\sum_{n }\frac{1}{\sqrt{n}}
\bigg[\(\an e^{-n\(t-ix \)}
+ \an^{\dagger }e^{n\(t -ix\)}\)+\({\bar a}_{n}^{I} e^{-n(t+ix)}
+{\bar a}_{n}^{\dagger }
e^{n(t + ix)}\)\bigg] \ ,
\end{equation}
and
\begin{equation}
\widetilde{\phi}\(t,x\)=
\sum_{n }\frac{1}{\sqrt{n}}
\bigg[\(\ta e^{-n\(\tilde{t}+i\tilde{x}\)}
+\ta^{\dagger }e^{n\(\tilde{t}+i\tilde{x}\)}\)
+\(\tab^{I}e^{-n\(\tilde{t}-i\tilde{x}\)}
+\tab^{\dagger\: I}e^{n\(\tilde{t}-i\tilde{x}\)}\)\bigg] \ ,
\end{equation}
where the expansion for $\widetilde{\phi}\(t,x\)$ is obtained from the tilde conjugation rules and $x\in[0,2\pi]$. The oscillators $a_n$ and $\bar{a}_n$ represent the holomorphic and anti-holomorphic sectors, respectively. The identification (\ref{ident}) turns out to be the following set of operatorial equations for a boundary state $|\phi\rangle=|\phi,\widetilde{\phi }\rangle$:
\begin{eqnarray}
\left[a_{n}^{I}-e^{-n\(\frac{\b}{2}+i\lambda\pi\)}\widetilde{a}^{\dagger\: I}_{n}\right]|\phi\rangle&=&0 \ , \label{bond1} \\
\left[\widetilde{a}_{n}^{I}-e^{-n\(\frac{\b}{2}+i\lambda\pi\)}{a}^{\dagger\: I}_{n}\right]|\phi \rangle&=&0 \ , \label{bond2} \\
\left[\bar{a}_{n}^{I}-e^{-n\(\frac{\b}{2}-i\lambda\pi\)}\widetilde{\bar{a}}^{\dagger\: I}_{n}\right]|\phi\rangle&=&0 \ , \label{bond3} \\
\left[\widetilde{\bar{a}}_{n}^{I}-e^{-n\(\frac{\b}{2}-i\lambda\pi\)}{\bar{a}}^{\dagger\: I}_{n}\right]|\phi\rangle &=&0 \ , \label{bond4}
\end{eqnarray}
with
\begin{equation}
|\Phi(q) \rangle=\left[\(q\bar{q}\)^\frac{1}{24}|\eta(q)|^{-2}\right]^{-\frac{1}{2}}e^{\sum_{n>0}\(a_{n}^{\dagger}\cdot {\tilde a}_{n}^{\dagger}\)q^\frac{n}{2}}\times e^{\sum_{n>0}\(\bar{a}_{n}^{\dagger}\cdot{\tilde{\bar a}}_{n}^{\dagger}\)\bar{q}^\frac{n}{2}}\left.\left|0\right\rangle\!\right\rangle \ , \label{bounds}
\end{equation}
where $q=e^{2\pi i\tau}$, $\eta\(q\)$ is the Dedekind $\eta$ function:
\begin{equation}
\eta(q)=q^{\frac{1}{24}}\prod_{n=1}\(1-q^n\) \ ,
\end{equation}
and $\bar{q}$ is the complex conjugate of $q$. We can verify that the expected values obtained in the state written above correspond to the following statistical average:
\begin{equation}
\left\langle O\right\rangle=\frac{{\mbox Tr}[O e^{-\b H'}]}{Z'} \ ,
\end{equation}
where the partition function $Z'$ is the oscillator part of the torus partition function 
\begin{equation}
Z'=Tre^{2\pi Im(\tau)H'}e^{2\pi Re(\tau)P'} \ ,
\end{equation}
and the operators $H'$ and $P'$ are the time and space translation operators on the cylinder, which are written in terms of holomorphic and anti-holomorphic oscillator number operators ($N_k$ and $\bar{N}_k$):
\begin{equation}
H'=\sum_{k=1}^{\infty}\left(N_k+\bar{N}_k\right)-\frac{(c+\bar{c})}{24} \ \ \ , \ \ \ P'=\sum_{k=1}^{\infty}\left(N_k-\bar{N}_k\right) \ .
\end{equation} 
The central charge $c=\bar{c}$ does not contribute to the statistical average. The partition function can be written as
\begin{equation}
Z' = Tr\left(q^{L_0-\frac{1}{24}}\bar {q}^{\bar{L}_0-\frac{1}{24}}\right)=|\eta(q)|^{-2} \ . 
\end{equation}
Note that $Z'$ is not modular invariant: this is because we do not take into account the continuous degrees of freedom of the zero mode, which are shared by the left and right movers. The zero mode wave function is $e^{ipx}$ and it has energy $L_0$, $\bar {L}_0=\frac{p^2}{2}$.  The zero mode partition function is $Z_0=\int dp q^{\frac{p^2}{2}}\bar{q}^{\frac{p^2}{2}}$ and the total partition function, $Z=Z'Z_0$, is modular invariant.  To incorporate the zero mode part into the TFD torus state, we just need to write the state as
\begin{eqnarray}
|\phi(p)\rangle &=& |\phi\rangle\otimes\Psi(\b,p) \ , \nonumber \\
\langle\phi(p)| &=& \langle\phi|\otimes\bar{\Psi}(\b,p) \ ,
\end{eqnarray}
where $\Psi(\b,p)$ is the normalized zero mode TFD momentum wave function:
\begin{eqnarray}
\Psi(q,p)&=& \langle p|0(\beta)\rangle = (Im(\tau))^{\frac{1}{4}}q^{\frac{p^2}{2}} \ , \nonumber \\
\bar{\Psi}(q,p)&=& \langle 0(\beta)|p\rangle = (Im(\tau))^{\frac{1}{4}}\bar{q}^{\frac{p^2}{2}} \ .
\end{eqnarray}
The zero mode dependence of the torus state will not be important for the entropy calculation we intend to make.

Note that the torus boundary state is an entanglement state involving $\phi$ and $\widetilde{\phi}$. Now, let us define the torus boundary state for a more general moduli. Assume that the moduli space confines the system to a region $R_1=[0,\Sigma]$. We also define the complement in the region $R_2=[\Sigma,\Lambda]$, where $\Lambda$ is an infrared cutoff. Our strategy is to calculate the entanglement entropy produced by the entanglement of the degrees of freedom defined in $R_1$ with the degrees of freedom defined in $R_2$, through the expected value of an operator written in terms of the degrees of freedom contained in $R_1$ - this operator is the entropy operator that defines the torus boundary state in an appropriate moduli. This state can be written as
\begin{equation}
|\phi\rangle=e^{-K}e^{\Sigma_n\(\widetilde{a}_{n}^{\dagger}a_{n}^{\dagger}+\widetilde{\bar{a}}_{n}^{\dagger}\bar{a}_{n}^{\dagger}\)}\left|0\right\rangle \ , \label{boundary}
\end{equation}
where the operator $K$ is defined by
\begin{eqnarray}
K&=&-\sum_{n=1}\bigg\{a_{n}^{\dagger}\cdot a_{n}\ln\left(\sinh^{2}\left(\theta _{n}\right)\right)-a_{n}\cdot a_{n}^{\dagger}\ln\left(\cosh^{2}\left(\theta_{n}\right)\right)\bigg\} \nonumber\\
&&-\sum_{n=1}\left\{\bar{a}_{n}^{\dagger}\cdot\bar{a}_{n}\ln\left(\sinh^{2}\left(\bar{\theta }_{n}\right)\right)-\bar{a}_{n}\cdot\bar{a}_{n}^{\dagger}\ln\left(\cosh^{2}\left(\bar{\theta }_{n}\right)\right)\right\} \ , \label{k}
\end{eqnarray}
with the identifications
\begin{equation}
\tanh\(\theta_n\)=\bar{q}^{\frac{n}{2}} \ ,
\qquad
\tanh\(\bar{\theta}_n\)=q^{\frac{n}{2}} \ , \label{tan}
\end{equation}
\begin{equation}
\sinh^{2}\(\theta_{n}\)=\frac{1}{q^{n}-1} \ ,
\qquad
\sinh^{2}\( \bar{\theta}_{n}\)=\frac{1}{\bar{q}^{n}-1}. \label{distri}
\end{equation}
We are going to show that, for an appropriate choice of the moduli, the $K$ operator corresponds to the entanglement entropy operator and expected values taken over the boundary state will correspond to the trace over the $R_2$ region. Before doing this, we define a sequence of conformal transformations, the same used in \cite{wil}. First, from $\varsigma=x+it$, we go to $w$ by
\begin{equation}
w=-\frac{\sin\frac{\pi}{\Lambda}\left(\varsigma-\Sigma\right)}{\sin\frac{\pi}{\Lambda}\left(\varsigma\right)} \ .
\end{equation}
This places the degrees of freedom set in $R_1$ on the positive side of the real axis. Now, we can go from $w$ to $z$ by the usual relation:
\begin{equation}
z=\frac{1}{k}\ln w \ .
\end{equation}
The system is confined to a strip of width $\pi/k$ and length $L$. If we define the ultraviolet cutoff $\epsilon$ such that $\Sigma\rightarrow\Sigma+\epsilon$, we obtain
\begin{equation}
L=\frac{2}{k}\ln\(\frac{\Sigma}{\epsilon}\) \ . \label{seg}
\end{equation}
Now we have a torus with periods $\frac{2\pi}{k}$ and $L$. We can define the moduli as $\tau=\frac{2\pi}{kL}$, so that $q=\bar{q}=e^{\frac{-4\pi^2}{kL}}$. The expected value of $K$ in $|\Phi\rangle$ is then given by
\begin{equation}
S=2\sum_n\left[\(1+\frac{q^n}{1-q^n}\)\ln\(1+\frac{q^n}{1-q^n}\)-\frac{q^n}{1-q^n}\ln\(\frac{q^n}{1-q^n}\)\right] \ . \label{entropy}
\end{equation}

The moduli choice allows one to evaluate the sums in the continuous limit $(|\tau|<<1)$ and, given the definition of $q$, the entropy is written in terms of known integrals. Assuming that $\Sigma>>\Lambda$, the result for the entropy is
\begin{equation}
S=\frac{1}{3}\ln\(\frac{\Sigma}{\epsilon}\) \ , \label{resfinal}
\end{equation}
which is the well-known result for the entanglement entropy of the $c=1$ conformal theory defined in the segment $\Sigma$. 

Now, let us define
\begin{equation}
{\cal V}_{n}(\t)=\prod_{k}\left(\frac{\sinh\left(\theta_{k}\right)^{2n_{k}}}{\cosh\left(\theta_{k}\right)^{2n_{k}+2}}\right) \ ,
\end{equation}
with the identifications (\ref{distri}). Considering
\begin{equation}
\rho = \left|\Phi(q) \right\rangle \left\langle \Phi(q)\right| \ ,
\end{equation}
the reduced density operator is written as
\begin{eqnarray}
\rho_{A} &=& Tr_{B}\left[ \left|\Phi(q) \right\rangle \left\langle \Phi(q)\right| \right]
\nonumber
\\
&=& \sum_{n} {\cal V}_{n}(t)\left| n \right\rangle \left\langle n \right| \ .
\end{eqnarray}
By tracing over the $B$-degrees of freedom (corresponding to $\widetilde{\phi}$), we obtain 
\begin{equation}
\rho_A (q) = e^{-{\cal S}_A (q)} \ ,
\end{equation}
which shows that the entropy operator ${\cal S}_A (q)$ is nothing but the so-called modular hamiltonian for this state \cite{haag}.

%%%%%%%%%%%%%%%%%%%%%%%%%%%%%%%%%%%%%%%%%%%%%%%%%%%%%%%%%%%%%%%%%%%%%

\section{The Neumann Boundary conditions} \label{Neumann}

This section aims to address the question of how to incorporate the boundary terms contribution  to the entanglement entropy calculation from the entropy operator. We are  going to find  contributions to the entanglement entropy from the choice of Neumann boundary conditions, which can  be indentified with the boundary entropy of the associated state in boundary conformal field theory \cite{Cardy:1989vyr}. We are going to work with a compact boson confined in the same region $L$ defined before.

The field expansion for a compact boson ($\Phi(x,t) \sim\ \Phi(x,t)+ \phi$) with Neumann boundary conditions is given by: 
\be
\Phi(x,t)&=&\Phi_0+\Pi_0 t+\sum_{k=1}^\infty \Phi_k(x,t) \ , \nonumber \\
\Phi_k(x,t)&=&{1\over\sqrt{\pi k}}\cos(\w_{k}x)\bigl[a_k e^{-i\w_k t}+\ad_k e^{i\w_k t}\bigr] \ , \nonumber \\
\Phi_0 &\sim& \Phi_0+\phi \ ,
\ee
with $\w_k = \pi k/L$. The canonical momentum is defined by $\Pi(x,t)=\partial_t\Phi(x,t)$ and the commutation relations obeyed by the fields and by the operators $a_k$ and $a_{k}^{\dag}$ are written as    
\be
[a_k,\ad_{k'}]=\delta_{kk'} \ \ \ , \ \ \  [\Phi(x,t),\Pi(x',t)]=i\delta(x-x') \ . \label{ccr}
\ee
The field obeys the following boundary conditions:
\be
\partial_x\Phi(x,t)\big|_{x=0} = 0 \quad  \ , \ \quad \partial_x\Phi(x,t)\big|_{x=L} = 0 \ .
\label{nbc}
\ee
The  zero modes are then given by
\be
\Phi_0=\int_0^L dx\;\Phi(x,t) \quad \ , \ \quad \Pi_0=\int_0^L dx\;\partial_t\Phi(x,t) \ .
\label{ph0pi0}
\ee
In order to use the same strategy previously defined, we duplicate the Hilbert space by defining the tilde fields $\tilde{\Phi}(x,t) = J\Phi(x,t)J $, and confine the system in a time interval $\beta$  imposing  the following gluing conditions:
\begin{eqnarray}
&&\Phi_k(t,x)-\widetilde{\Phi}_k(-\tilde{t}-\frac{\b}{2},x)=0 \ , \nonumber\\
&&\Phi_0=\widetilde{\Phi_0} - \frac{\b}{2}\widetilde{\Pi_0} \ , \nonumber \\
&&\Pi_0=\widetilde{\Pi_0} \ . 
\end{eqnarray}
 
The boundary state is now defined by
\be
|\Phi\rangle=|\Phi\rangle_{OS} \ |\Phi\rangle_0 \ , 
\ee
where $|\Phi\rangle_{OS}$ is the oscillator part and $|\Phi\rangle_0$ is the zero mode part. The normalized oscillator contribution is given by
\be
|\Phi\rangle_{OS}=\bigl[e^{\pi\beta/24L}\, \eta(q)\bigr] \ e^{\sum \ad_k\widetilde{a}^{\dagger}_k q^k}|0\rangle\widetilde{|0\rangle} \ , 
\ee
with $q=e^{-\frac{\beta \pi}{L}}$. Now we have an annulus of length $L$ and circumference $\beta$ and this boundary state is analogous to a thermal state of a one-dimensional QFT in an interval of length $L$, at the temperature $\beta^{-1}$.

First we will focus on the oscillator part. One can use the same parameters defined by the previous conformal transformations, $\beta= 2\pi/k $ and $L$, given by Eq.~(\ref{seg}). The entropy operator is the same as before, with $q= e^{-\frac{2\pi^2}{k L}}$, and the oscillator contribution for the entropy is the same as in (\ref{entropy}), without the factor of $2$, which comes from the holomorphic and anti-holomorphic contributions. However, the final result is the same as in Eq.~(\ref{resfinal}), just because a factor of $2$ appears in the new $q$ parameter. 
 
%%%%%%%%%%%%%%%%%%%%%%%%%%%%%%%%%%%%%%%%%%%%%%%%%%%%%%%%%%%%%%%%%%% 
   
\subsection{The zero mode} \label{zero}

For the zero mode, we do not have, at first, an oscillator Heisenberg  algebra. So, the boundary state do not have the same structure as before. However, we can generalize the method by using an analogy with the TFD thermal state. From Eqs.~(\ref{ccr}) and (\ref{ph0pi0}), the zero mode satisfies the canonical commutation relation:
\be
[\Phi_0,\Pi_0]={i\over L} \ . \label{ccr0}
\ee

Since $\Phi_0$ is compact ($\Phi_0 \sim \Phi_0+\phi$), the zero mode canonical momentum has discretized spectrum. In fact, the zero mode wave function is given by $\Psi[\Phi_0] \approx e^{i\Pi_0 \Phi_0 L}$, where $\Pi_0 $ can be written as 
\be
\Pi_0 = {2\pi\over L\phi} \ \hat{P} \ ,
\ee
and the operator $\hat{P}$ has the following eigenfunctions:
\be
\hat P|m\rangle=m|m\rangle,\quad m=0,\pm1,\ldots
\ee

The zero mode hamiltonian is given by
\begin{equation}
H_0= {2\pi^2\over L\Phi^2}\hat{P}^2 \ .
\end{equation}
Thus, using the basis $|m\rangle$, the zero mode thermal state, which is a solution of Eq.~(\ref{tscg}), can be written as  
\be
|\Phi\rangle_0  =\frac{1}{\vartheta(i\tau,0)}\sum_{m=-\infty }^{m=+\infty}e^{-\frac{\beta}{2}{2\pi^2\over L\Phi^2}\hat{P}^2}|m\rangle \ ,
\ee
with $\tau =2\pi \beta/L\Phi^2$, and the Jacobi Theta function is defined by 
\be
\vartheta(i\tau,0)=\sum_{m=-\infty}^{+\infty}e^{i\pi\tau m^2} \ .
\ee
Now we can use the same $\beta$ and $L$ as before and interpret the state just as a general entanglement state, and not as a thermal one. The state can be generated again by an entropy operator, as in Eq.~(\ref{boundary}):
\be
|\Phi\rangle_0 = e^{-S_0}I \ \ \ , \ \ \ I=\sum|m\rangle\widetilde{|m\rangle} \ ,
\ee
where the zero mode entropy operator has the same structure defined before and can be written as 
\be
S_0= \hat{P}^2\ln q^{-\tau}-\ln \vartheta(i\tau,0) \ ,
\ee
which has the same form as in (\ref{entgeral}). To evaluate the expected value of $S_0$ in $|BN\rangle_0$  we need just the result
\be
{}_{0}\langle\Phi|\hat{P}^2|\Phi\rangle_0=-i\frac{d}{d\tau}\ln \vartheta(i\tau,0) \ .
\ee
By using the modular properties of the Theta function, $\vartheta(\tau,0)=(-i\tau)^{-1/2}\vartheta(-1/\tau,0)$, and the product representation
\be
\vartheta(\tau,0)=\prod_{n=1}\left(1-e^{n2\pi i\tau}\right)\left(1+e^{(n-1/2)(2\pi i\tau)}\right)^2 \ ,
\ee
the expected value of $S_0$  for large $L$ is given by
\be
K_0=\langle 0,\tau|S|0,\tau\rangle= {1\over 2}\log\!\left({2L\over \beta}\right)
+{1\over 2}\log\!\left({\phi^2\over 4\pi}\right)+... 
\ee 
The first term is an anomalous and subleading term $\log\!\left({2L\over \beta}\right)$. This term has been remarked previously in the literature \cite{Gutperle:2017enx}, where it was derived from the contribution to entropy due to conformal junctions. The second term, independent of $L$, can be understood as a contribution from the Affleck-Ludwig boundary entropy \cite{Affleck} and was derived also in \cite{Michel:2016fex}. The dots represent the terms of order $1/L$.

%%%%%%%%%%%%%%%%%%%%%%%%%%%%%%%%%%%%%%%%%%%%%%%%%%%%%%%%%%%%%%%%%%%%

\section{A dissipative model for entanglement evolution} \label{dissipative}

Once we have shown that we can use the TFD canonical approach to calculate the entanglement entropy, let us propose a model for the entanglement evolution. Inspired in the non-equilibrium TFD formulation, we propose a local interaction between the system and the auxiliary one. The simplest model is given by the hamiltonian:
\begin{equation}
H_I= i\gamma \sum_n \left(a_n\widetilde{a}_n-a_n^{\dagger}\widetilde{a}^{\dagger}_n + \bar{a}_n\widetilde{\bar{a}}_n-\bar{a}_n^{\dagger}\widetilde{\bar{a}}^{\dagger}_n\right) \ .
\end{equation}
Suppose that this interaction is turned on, at least for a short period of time. This is a typical dissipative hamiltonian, explored a long time ago in canonical dissipation theory \cite{ceravi}; the $\lambda$ parameter is the damping term. This hamiltonian produces an exchange of energy between the system and its copy. 

The total hamiltonian is written as: 
\begin{equation}
H=\sum_n\left(a_na^{\dagger}_n-\widetilde{a}_n\widetilde{a}^{\dagger}_n\right)+i\gamma\left(a_n\widetilde{a}_n-a_n^{\dagger}\widetilde{a}^{\dagger}_n\right)+\sum_n\left(\bar{a}_n\bar{a}^{\dagger}_n-\widetilde{\bar{a}}_n\widetilde{\bar{a}}^{\dagger}_n\right)+i\gamma\left(\bar{a}_n\widetilde{\bar{a}}_n-\bar{a}_n^{\dagger}\widetilde{\bar{a}}^{\dagger}_n\right) \ .
\end{equation}
Note that the hamiltonian is invariant under tilde conjugation rules. The idea is to evolve the state defined in (\ref{boundary}) in time, using this hamiltonian. The time-dependent state becomes:
\begin{eqnarray}
\left|\Phi(q,t)\right\rangle&=&\exp(-iHt)\left|0(t=0)\right\rangle \nonumber\\
&=&\exp\left[\sum_{k}\; (\frac{\gamma t}{2}+\theta_k)\left(A_{-k}B_{k}-A^{\dagger}_{-k}B^{\dagger}_k\right)\right]|0\rangle_A\otimes|0\rangle_B\nonumber\\
&=&\prod_{k}\frac{\delta_{kk}}{\cosh\left(\frac{\gamma t}{2}+\theta_k\right)}\exp\left[\tanh\left(\frac{\gamma t}{2}+\theta_k\right)A^{\dagger}_{-k}B^{\dagger}_{k}\right]|0\rangle_A\otimes|0\rangle_B \ , \label{res1}
\end{eqnarray}
where
\begin{equation}
\theta_n=\frac{1}{2}\ln\left[\frac{1+q^{\frac{n}{2}}}{1-q^{\frac{n}{2}}}\right] \ . \label{thetan}
\end{equation}

The expected value of the number operator is given by
\begin{equation}
N_k(q,t)=\langle \Phi(q,t)|a_k^{\dagger}a_k|\Phi(q,t)\rangle=\sinh^2(\theta_n+\gamma t) \ .
\end{equation} 
Therefore, we have a simple way to obtain the temporal evolution of the entropy. First we need to show that the time-dependent entanglement state ($\ref{res1}$) can also be generated by a time-dependent entropy operator, in order to get the same structure of Eq.~(\ref{boundary}). The obvious generalization is
 \begin{eqnarray}
K(q,t)&=&-\sum_{n=1}\bigg\{a_{n}^{\dagger}\cdot a_{n}\ln\left(\sinh^{2}\left(\gamma t+\theta _{n}\right)\right)-a_{n}\cdot a_{n}^{\dagger}\ln\left(\cosh^{2}\left(\gamma t+ \theta_{n}\right)\right)\bigg\} \nonumber\\
&&-\sum_{n=1}\left\{\bar{a}_{n}^{\dagger}\cdot\bar{a}_{n}\ln\left(\sinh^{2}\left(\gamma t+\bar{\theta }_{n}\right)\right)-\bar{a}_{n}\cdot\bar{a}_{n}^{\dagger}\ln\left(\cosh^{2}\left(\gamma t +\bar{\theta }_{n}\right)\right)\right\} \ , \label{k}
\end{eqnarray}
with $\theta_n$ given by (\ref{thetan}).  It easy to show that 
\be
\left|\Phi(q,t)\right\rangle= e^{-\frac{1}{2}K}e^{\Sigma_n\(\widetilde{a}_{n}^{\dagger}a_{n}^{\dagger}+\widetilde{\bar{a}}_{n}^{\dagger}\bar{a}_{n}^{\dagger}\)}\left|0\right\rangle \ , 
\ee
such that the same structure defined for the non-dynamical case is maintained. Now we just need to evaluate the expected value of the operator (\ref{k}) in the state (\ref{res1}). The time-dependent entanglement entropy is 
\begin{equation}
S_A(q,t)=\sum_{k}\left[(1+N_k(q,t))\ln(1+N_k(q,t))-N_k(q,t)\ln N_k(q,t)\right] \ , \label{Sphys} 
\end{equation}
where, as previously defined, $q=\bar{q}=e^{\frac{-4\pi^2}{kL}} $, and $L$ is given in (\ref{seg}). The time-dependent entropy for $\gamma t << 1$ is given by
\begin{eqnarray}
S&=&\frac{1}{3}\ln\(\frac{\Sigma}{\epsilon}\)+\gamma t\sum\frac{q^\frac{n}{2}\ln(q^n)}{q^n-1} \nonumber\\
&=&\frac{1}{3}\ln\(\frac{\Sigma}{\epsilon}\)-\gamma t\frac{\pi^2}{2\ln q} \nonumber\\
&=& \ln\(\frac{\Sigma}{\epsilon}\) \left(\frac{1}{3}+\frac{\gamma t}{4}\right) \ . \label{Sentanglement}
\end{eqnarray}
This result reproduces the well-known linear growth of the entanglement entropy in conformal theories \cite{calabrese}. Note that the time-dependent part is also proportional to $\ln\(\frac{\Sigma}{\epsilon}\)$, as one should expect.

The time-dependent entanglement state $(\ref{res1}) $ has a rich interesting behavior at later times. Note that
\be
\left\langle \Phi(q,t)|\Phi(q,t) \right\rangle &=&1 \ , \forall t \ , \nonumber \\
\left\langle \Phi(q,t=0)|\Phi(q,t) \right\rangle &=& e^{-\sum_n\cosh(\theta_n+\gamma t)} \ .
\ee
So, at later times we obtain:
\be
\lim_{t\rightarrow\infty}\left\langle \Phi(q,t=0)|\Phi(q,t) \right\rangle=0 \ .
\ee
This is a characteristic of  non-unitary time evolution, which is in general observed in entanglement states \cite{Brustein:2006wp}. The vacuum state at $t= 0$ and the vacuum at later times are orthogonal and the corresponding Hilbert spaces are unitarily inequivalent representations of the canonical commutation relations. From the GNS point of view, each vacuum induce a different representation of the operator algebra.  These features also reflect indeed correctly the irreversibility of the time evolution characteristic of dissipative theories. In fact, the time-dependent entanglement state satisfies:
\be
\frac{\partial \left|\Phi(q,t)\right\rangle}{\partial t}= -\frac{1}{2}\left(\frac{\partial S(q,t)}{\partial t}\right)\left|\Phi(q,t)\right\rangle \ .
\ee
Thus, the entanglement dynamics is controlled by the variation of the entanglement entropy operator. 
The connection between dissipation and entanglement is not surprising and was demonstrated experimentally in  \cite{EntanglementDissipation}. It was explored within the holographic context in \cite{BGMN}. However, the entanglement entropy calculated in \cite{BGMN} is the finite temperature momentum entanglement an is not related to a spatial partitioned Hilbert space, as the entropy calculated here with the techniques developed in the present work.  

%%%%%%%%%%%%%%%%%%%%%%%%%%%%%%%%%%%%%%%%%%%%%%%%%%%%%%%%%%%

\section{Conclusions} \label{conc}

A rigorous computation of the entanglement entropy requires some methodology which allows to compute it directly, without use of the replica trick and of the analytical continuation from R\'enyi entropy. In this work, such a methodology is developed, by defining a canonical way, without euclidean path integral manipulations, to calculate the entanglement entropy for $c=1$ conformal theories. We reproduce results for bosonic fields with periodic and Neumann boundary conditions. For the later, the zero mode plays an important role.

Since the usual method is based on an analogy with the SK formulation of real time finite temperature field theory, here we developed a method based on an analogy with the TFD canonical formulation. Although the integral path TFD perturbation theory coincides with the SK formulation, there are conceptual differences between these formulations that make TFD more suitable to be used in the construction of a method that allows to calculate entropy outside the thermodynamic context. The central issue is the duplication of the degrees of freedom. While in the SK formulation the duplication is a consequence of the contour and so the fields represent the algebra of operators in ordinary Fock space, in TFD the doublet fields are related to the GSN representation induced by KMS states. Thus, it is more natural to generalize the TFD technology to a construction that takes into account more general mixed states. 

The starting point of the methodology developed here is the definition of the thermal vacuum. We have shown that it is possible to generalize the notion of thermal vacuum to a boundary state which confines the fields in a general temporal segment $\beta$, which is not necessarily related to temperature. The boundary state is related to a torus for fields with periodic boundary conditions and to an anulus for fields with Neumann boundary conditions. From the point of view of the GNS representation, the boundary state is a more general cyclic and separating vector. We have shown that the choice of the moduli space is related to the parameters of the Bogoliubov transformation used in TFD.

Even more interesting, the TFD entropy operator translates to an entanglement entropy operator and we have shown that its expected value provides the entanglement entropy between the degrees of freedom defined in a line segment and its complement. Note that, in general, boundary states essentially have no space entanglement \cite{Masamichi}. Here we have constructed a boundary state that has real space entanglement entropy. The canonical formalism developed here allows to construct a simple model for entropy evolution and the typical linear behavior of conformal theories is achieved. At large times, the entanglement vacuum is not unitarily equivalent to the initial state, which is typical of entangled systems. The non-unitary dynamics is related to the fact that, at large times, the vacuum induces a truly different representation of the operator algebra. Note that from the theorem described by Eq.~(\ref{TTT}), we can conclude that in the modular theory the notion of time depends on the state, which  suggests a kind of parallelism with general relativity. Thus, the fact that the time evolution of the entanglement state is generated by the time variation of the entanglement operator shows that we can try to make some connection of the framework developed here with quantum gravity. The key to do such endeavour is to use the Ryu-Takayanagi formula to   relate the kind of entropy operator formulated here with the area operator \cite{kamal}. In fact, the canonical method presented here can be further developed in order to give issues for an algebraic program to calculate entanglement entropy in the holographic context. A good start point is to reproduce the results of \cite{Harlow:2016vwg} involving quantum error correcting code. Also, the dictionary between subalgebras and boundary conditions in the path integral presented in \cite{Lin:2018bud} should be revisited in the canonical approach. The canonical formalism also makes it easier to calculate the entropy quantum fluctuation, as studied in \cite{quantfluc}.

Finally, we have presented in this work a methodology that allows to calculate the conformal field entanglement entropy in such a way that it is possible to draw some correspondences with the algebraic approach. In fact, TFD establishes a dictionary between a more formal and abstract approach and an approach written in a more physical language. We show that the TFD entropy operator can be  related to the modular hamiltonian. It is well-known that an algebraic von Neumann entropy admits no definition for general quantum systems, but related quantities, such as relative entropy, is algebraically well-defined. Thus, it is very important to connect the entanglement operator defined here with the Araki's algebraic relative entropy \cite{Araki:1976zv}, which is indeed written in terms of the modular hamiltonian.

%%%%%%%%%%%%%%%%%%%%%%%%%%%%%%%%%%%%%%%%%%%%%%%%%%%%%%%%%%%%%%%%%%%%%%%%%%%%%%%%%%%%%%%%%%%%%%%%%%%%%%%%%%%%%%%%%%%%%%%%%%%%%%%%%%%%%%%%%

\begin{acknowledgments}
The authors would like to thank D. F. Z. Marchioro for useful discussions.
\end{acknowledgments}

%%%%%%%%%%%%%%%%%%%%%%%%%%%%%%%%%%%%%%%%%%%%%%%%%%%%%%%%%%%%%%%%%%%%%%%%%%%%%%%%%%%%%%%%%%%%%%%%%%%%%%%%%%%%%%%%%%%%%%%%%%%%%%%%%%%%%%%%%

%%%%%%%%%%%%%%%%%%%%%%%%%%%%%%%%%%%%%%%%%%%%%%%%%%%%%%%%%%%%%%%%%%%%%%%%%%%%%%%%%%%%%%%%%%%%%%%%%%%%%%%%%%%%%%%%%%%%%%%%%%%%%%%%%%%%%%


\begin{thebibliography}{99} 

\bibitem{app} 
P.~Benioff, 
Phys.\ Rev.\ Lett. {\bf48} (1982) 1581.

W.~K.~Wooters and W.~H.~Zurek, 
Nature {\bf299} (1996) 802.

\bibitem{Kitaev}
A. Kitaev and J. Preskill, 
%Topological entanglement entropy, 
Phys.\ Rev.\ Lett.\ {\bf96} (2006) 110404, 
[arXiv:hep-th/0510092].

\bibitem{Levin} 
M.~Levin and X.-G.~Wen, 
%Detecting Topological Order in a Ground State Wave Function,
Phys.\ Rev.\ Lett.\ {\bf96} (2006) 110405, 
[arXiv:cond-mat/0510613].

\bibitem{Maldacena} 
J.~M.~Maldacena, 
%The Large N limit of superconformal Field theories and supergravity, 
Int.\ J.\ Theor.\ Phys.\ {\bf38} (1999) 1113, 
Adv.\ Theor.\ Math.\ Phys.\ {\bf2} (1998) 231,
[arXiv:hep-th/9711200].

\bibitem{RT}
S.~Ryu and T.~Takayanagi, 
%Holographic derivation of entanglement entropy from AdS/CFT, 
Phys.\ Rev.\ Lett.\ {\bf96} (2006) 181602, 
[arXiv:hep-th/0603001].

\bibitem{takareview}
M.~Rangamani and T.~Takayanagi,
%Holographic Entanglement Entropy,
Lec.\ Notes\ Phys.\ {\bf931} (2017) 1, 
[arXiv:1609.01287 [hep-th]].

\bibitem{SK} 
N.~P.~Landsman and C.~G.~van Weert,
%``Real and Imaginary Time Field Theory at Finite Temperature and Density,''
Phys.\ Rept.\  {\bf 145} (1987) 141.

\bibitem{witten} 
E.~Witten,
  %``APS Medal for Exceptional Achievement in Research: Invited article on entanglement properties of quantum field theory,''
Rev.\ Mod.\ Phys.\  {\bf 90} (2018) no.4, 045003,
[arXiv:1803.04993 [hep-th]].

\bibitem{Reznik:1995hy}
B.~Reznik,
%``Unitary evolution between pure and mixed states,''
Phys.\ Rev.\ Lett.\ \textbf{76} (1996) 1192,
%doi:10.1103/PhysRevLett.76.1192
[arXiv:quant-ph/9504019].

\bibitem{Banks:1983by}
T.~Banks, L.~Susskind and M.~E.~Peskin,
%``Difficulties for the Evolution of Pure States Into Mixed States,''
Nucl.\ Phys.\ B \textbf{244} (1984) 125.
%doi:10.1016/0550-3213(84)90184-6

\bibitem{Brustein:2006wp}
R.~Brustein, M.~B.~Einhorn and A.~Yarom,
%``Entanglement and Nonunitary Evolution,''
JHEP \textbf{04} (2007) 086,
%doi:10.1088/1126-6708/2007/04/086
[arXiv:hep-th/0609075].

\bibitem{ume2}
Y.~Takahashi and H.~Umezawa,
Coll.\ Phenomena {\bf 2} (1975) 55
(Reprinted in Int.\ J.\ Mod.\ Phys.\ B {\bf 10} (1996) 1755).

\bibitem{ume4}
H.~Umezawa, H.~Matsumoto, M.~Tachiki,
{\it Thermofield Dynamics and Condensed States}
(North-Holland, Amsterdan, 1982).

\bibitem{KKMS} 
F.~C.~Khanna, A.~P.~C.~Malbouisson, J.~M.~C.~Malbouisson and A.~E.~Santana,
{\it Thermal Quantum Field Theory - Algebraic Aspects and Applications}
(World Scientific, Singapore, 2009).

\bibitem{Arimitsu}	
T.~Arimitsu, J.~Phradko and H.~Umezawa, 
Physica A {\bf 135} (1986) 487.
	
\bibitem{Marinaro} 
M.~Marinaro, 
Phys.\ Rept.\ {\bf 137} (1986) 81.

\bibitem{TT}
M.~Tomita, 
Kyushu University preprint (1967).

M.~Takesaki, 
{\it Tomita's Theory of Modular Hilbert Algebras and Its Applications, Lecture Notes in Mathematics, Vol. 128} (Springer-Verlag, Berlin, Heidelberg and New York, 1970).

\bibitem{ojima81}
I.~Ojima,
%``Gauge Fields at Finite Temperatures: Thermo Field Dynamics, KMS Condition and their Extension to Gauge Theories,''
Annals\ Phys.\  {\bf 137} (1981) 1.

\bibitem{kamal} 
H.~Kamal and G.~Penington,
%``The Ryu-Takayanagi Formula from Quantum Error Correction: An Algebraic Treatment of the Boundary CFT,''
[arXiv:1912.02240 [hep-th]].

\bibitem{Harlow:2016vwg}
D.~Harlow,
%``The Ryu\textendash{}Takayanagi Formula from Quantum Error Correction,''
Commun.\ Math.\ Phys.\ \textbf{354} no.3 (2017) 865, 
%doi:10.1007/s00220-017-2904-z
[arXiv:1607.03901 [hep-th]].

\bibitem{barata}
J.~C.~A.~Barata, M.~Brum, V.~Chabu and R.~Correia~da~Silva,
Braz.\ J.\  Phys.\ (2020),
%doi:10.1007/s13538-020-00808-0
[arXiv:1909.06232 [math-ph]].

\bibitem{kadiring}
R.~V.~Kadison and J.~R.~Ringrose, 
{\it Fundamentals of the Theory of Operator Algebras, Vol. I} 
(Academic, New York, 1983).

\bibitem{muw}
H.~Matsumoto, H.~Umezawa and J.~P.~Whitehead, 
Prog.\ Theor.\ Phys.\ {\bf 76} (1986) 260.

\bibitem{mat1}
H.~Matsumoto, 
{\it in: Progress in Quantum Field Theory, eds. H. Ezawa and S. Kamefuchi} 
(North-Holland, Amsterdam, 1986).

\bibitem{mat2}
H.~Matsumoto,
{\it in: Quantum Field Theory, ed. F. Mancini} 
(North-Holland, Amsterdam, 1986).

\bibitem{land85}
N.~P.~Landsman,
{\it Self energy and composite operators in real time finite temperature field theory},
Report number: ITFA-85-08 (1985).

\bibitem{Fewster:2019ixc}
C.~J.~Fewster and K.~Rejzner,
%``Algebraic Quantum Field Theory -- an introduction,''
[arXiv:1904.04051 [hep-th]].

\bibitem{GN}
I.~M.~Gelfand and M.~A.Naimark,
%On the imbedding of normed rings into the ring of operators on a Hilbert space, 
Matematicheskii\ Sbornik\ {\bf 12} (2) (1943) 197.

\bibitem{S}
I.~E.~Segal, 
%Irreducible representations of operator algebras, 
Bull.\ Amer.\ Math.\ Soc.\ {\bf 53} (1) (1947) 73.

\bibitem{nos}  
M.~Botta Cantcheff, A.~L.~Gadelha, D.~F.~Z.~Marchioro and D.~L.~Nedel,
%``Entanglement from Dissipation and Holographic Interpretation,''
Eur.\ Phys.\ J.\ C {\bf 78} (2018) no.2, 105,
[arXiv:1702.02069 [hep-th]].

\bibitem{Emch}
G.~G.~Emch, 
{\it Algebraic Methods in Statistical Mechanics and Quantum Field Theory} 
(Wiley \& Sons, New York, 1972).

\bibitem{braro}
O.~Bratteli and D.~W.~Robinson, 
{\it Operator Algebras and Quantum Statistical Mechanics, Vols. I and II} 
(Springer, New York, 1979, 1981).

\bibitem{hhw}
 R.~Haag, N.~M.~Hugenholtz and M.~Winnink, 
 Comm.\ Math.\ Phys.\ {\bf 5} (1967) 215.
 
\bibitem{ume1}
H.~Umezawa,
{\it Advanced Field Theory: Micro, Macro and Thermal Physics}
(AIP, New York, 1993).

\bibitem{ChuUme}
H.~Chu and H.~Umezawa,
%``A Unified formalism of thermal quantum field theory'',
Int.\ J.\ Mod.\ Phys.\ A {\bf 9} (1994) 2363.

\bibitem{Hen}
P.~A.~Henning,
%``Thermo field dynamics for quantum fields with continuous mass spectrum'',
Phys.\ Rept.\ {\bf 253} (1995) 235.

\bibitem{gadcris} 
M.~C.~B.~Abdalla and A.~L.~Gadelha,
%``General unitary SU(1,1) TFD formulation,''
Phys.\ Lett.\ A {\bf 322} (2004) 31, 
[arXiv:hep-th/0309254].

\bibitem{adsnos}  
M.~B.~Cantcheff, A.~L.~Gadelha, D.~F.~Z.~Marchioro and D.~L.~Nedel,
%``String in AdS Black Hole: A Thermo Field Dynamic Approach,''
Phys.\ Rev.\ D {\bf 86} (2012) 086006,
[arXiv:1205.3438 [hep-th]].

\bibitem{gad1} 
M.~C.~B.~Abdalla, A.~L.~Gadelha and I.~V.~Vancea,
%``On the SU(1,1) thermal group of bosonic strings and D-branes'',
Phys.\ Rev.\ D {\bf 66} (2002) 065005,
[arXiv:hep-th/0203222].

\bibitem{gad2} 
M.~C.~B.~Abdalla, A.~L.~Gadelha and D.~L.~Nedel,
%``On the entropy operator for the general SU(1,1) TFD formulation'',
Phys.\ Lett.\ A {\bf 334} (2005) 123,
[arXiv:hep-th/0409116].

\bibitem{Brustein}
Ram~Brustein, Martin~B.~Einhorn and Amos~Yarom,
%``Entanglement interpretation of black hole entropy in string theory'',  
JHEP {\bf 0601} (2006) 098,
[arXiv:hep-th/0508217].

\bibitem{torus}
M.~C.~B.~Abdalla, A.~L.~Gadelha and D.~L.~Nedel,
 %``Closed string thermal torus from thermofield dynamics,''
Phys.\ Lett.\ B {\bf 613} (2005) 213,
[arXiv:hep-th/0410068].

\bibitem{wil}  
C.~G.~Callan, Jr. and F.~Wilczek,
%``On geometric entropy,''
Phys.\ Lett.\ B {\bf 333} (1994) 55,
[arXiv:hep-th/9401072].

\bibitem{haag} 
R.~Haag,
{\it Local Quantum Physics: Fields, Particles,
Algebras}
(Springer-Verlag, New York, 1992).

\bibitem{Cardy:1989vyr}
J.~Cardy,
%``Boundary Conditions in Conformal Field Theory,''
Adv. Stud. Pure Math. \textbf{19} (1989) 127.

\bibitem{Gutperle:2017enx} 
M.~Gutperle and J.~D.~Miller,
%``Entanglement entropy at CFT junctions,''
Phys. Rev. D \textbf{95} (2017) no.10, 106008,
%doi:10.1103/PhysRevD.95.106008
[arXiv:1701.08856 [hep-th]].

\bibitem {Affleck} 
I.~Affleck and A.~W.~W.~Ludwig, 
%\emph{{Universal noninteger 'ground state degeneracy' in critical quantum systems}}, 
Phys. Rev. Lett. {\bf 67} (1991) 161.

\bibitem{Michel:2016fex} 
B.~Michel and M.~Srednicki,
%``Entanglement Entropy and Boundary Conditions in 1+1 Dimensions,''
[arXiv:1612.08682 [hep-th]].

\bibitem{ceravi}
E.~Celeghini, M.~Rasetti and G.~Vitiello,
%``Quantum dissipation,''
Annals\ Phys.\  {\bf 215} (1992) 156.

\bibitem{calabrese} 
P.~Calabrese and J.~L.~Cardy, 
%``Entanglement entropy and quantum field theory'', 
J.\ Stat.\ Mech.\ {\bf0406} (2004) P06002, 
[arXiv:hep-th/0405152].

\bibitem{EntanglementDissipation} 
H.~Krauter, C.~A.~Muschik, K.~Jensen, W.~Wasilewski, J.~M.~Petersen, J.~I.~Cirac and E.~S.~Polzik, 
%``Entanglement Generated by Dissipation and Steady State Entanglement of Two Macroscopic Objects'', 
Phys.\ Rev.\ Lett.\ {\bf 107} (2011) 080503, 
[arXiv:1006.4344 [quant-ph]].

\bibitem{BGMN}  
M.~Botta Cantcheff, A.~L.~Gadelha, D.~F.~Z.~Marchioro and D.~L.~Nedel,
%``Entanglement from Dissipation and Holographic Interpretation,''
Eur.\ Phys.\ J.\ C {\bf 78} (2018) no.2, 105,
[arXiv:1702.02069 [hep-th]].

\bibitem{Masamichi}
M.~Miyaji, S.~Ryu, T.~Takayanagi and X.~Wen,
%Boundary States as Holographic Duals of Trivial Spacetimes,
JHEP {\bf05} (2015) 152, 
[arXiv:hep-th/1412.6226].

\bibitem{Lin:2018bud}
J.~Lin and D.~Radi\v{c}evi\'c,
%``Comments on defining entanglement entropy,''
Nucl.\ Phys.\ B \textbf{958} (2020) 115118,
%doi:10.1016/j.nuclphysb.2020.115118
[arXiv:1808.05939 [hep-th]].

\bibitem{quantfluc}
Si-Xuan~Zhang, Tong-Hua~Liu, Shuo~Cao, Yu-Ting~Liu, Shuai-Bo~Geng and Yu-Jie~Lian,
%Quantum fluctuation of entanglement for accelerated two-level detectors,
Chin.\ Phys.\ B {\bf 29} (2020) 5, 050402,
[arXiv:2003.03783 [quant-ph]].

\bibitem{Araki:1976zv}
H.~Araki,
%``Relative Entropy of States of Von Neumann Algebras,''
Publ.\ Res.\ Inst.\ Math.\ Sci.\ Kyoto \textbf{1976} (1976), 809.


\end{thebibliography}
\end{document}